\documentclass{elsart}

\usepackage[centertags]{amsmath}
\usepackage{amssymb}
\usepackage[figuresright]{rotating}
\usepackage{newlfont}

\newcommand{\aut}{\mathop{\mathrm {Aut} }\nolimits}
\newcommand{\RE}{\mathop{\mathrm {Re} }\nolimits}
\newcommand{\IM}{\mathop{\mathrm {Im} }\nolimits}
\newcommand{\E}{{\cal E}}
\newcommand{\el}{{\cal L}}
\newcommand{\es}{{\cal S}}
\newcommand{\er}{{\cal R}}
\newcommand{\q}{\quad}
\newcommand{\pe}{{\cal P}}
\newcommand{\I}{{\cal I}}
\newcommand{\qe}{{\cal Q}}
\newcommand{\ep}{\varepsilon}
\newcommand{\N}{{\mathbb N}}

\newcommand{\Z}{\mathbb{Z}}
\def\C{\mathbb C}
\newcommand{\set}[1]{\left\{#1\right\}}

\newcommand{\p}{\cdot}
\newcommand{\bC}{\mathbb{C}}
\newcommand{\mL}{\mathcal{L}}
\newcommand{\mD}{\mathcal{D}}
\newcommand{\Der}{\operatorname{Der}}
\newcommand{\ad}{\operatorname{ad}}
\def\Z{\mathbb Z}

\begin{document}
\begin{frontmatter}
\title{Graded contractions of the Pauli graded
$sl(3,\C)$}

\author[Prague]{J.~Hrivn\'ak},
\author[Prague]{P.~Novotn\'y},
\author[Montreal]{J.~Patera} and
\author[Prague]{J.~Tolar}
\address[Prague]{Department of Physics,
Faculty of Nuclear sciences and Physical Engineering, Czech
Technical University, B\v{r}ehov\'a 7, 115 19 Prague 1, Czech
Republic}
\address[Montreal]{Centre de recherches math\'ematiques, Universit\'e de Montr\'eal, CP 6128, Succursale Centre-Ville, Montr\'eal, Canada H3C 3J7 }



\date{\today}
\begin{abstract}
The Lie algebra $sl(3,\C)$ is considered in the basis
of generalized Pauli matrices. Corresponding grading is
the Pauli grading here. It is one of the four gradings
of the algebra which cannot be further refined.

The set $\es$ of 48 contraction equations for 24 contraction
parameters is solved. Our main tools are the symmetry group of the
Pauli grading of $sl(3,\C)$, which is essentially the finite group
$SL(2,\Z_3)$, and the induced symmetry of the system $\es$. A list
of all equivalence classes of solutions of the contraction
equations is provided. Among the solutions, 175 equivalence
classes are non-parametric and 13 solutions depend on one or two
continuous parameters, providing a continuum of equivalence
classes and subsequently continuum of non-isomorphic Lie algebras.
Solutions of the contraction equations of Pauli graded $sl(3,\bC)$
are identified here as specific solvable Lie algebras of
dimensions up to 8. Earlier algorithms for identification of Lie
algebras, given by their structure constants, had to be made more
efficient in order to distinguish non-isomorphic Lie algebras
encountered here.

Resulting Lie algebras are summarized in tabular form. There are
88 indecomposable solvable Lie algebras of dimension 8, 77 of them
being nilpotent. There are 11 infinite sets of parametric Lie
algebra which still deserve further study.

\end{abstract}

\end{frontmatter}
\section{Introduction}

Study of homomorphic relations between pairs of Lie algebras, called algebra--subalgebra
pairs, over complex or real number field, was recognized as a naturally interesting and
useful problem with the advent of Lie theory over a century ago. During that century the
number and diversity of applications of Lie theory in mathematics, science, and
engineering was rapidly increasing. In parallel, it was becoming increasingly more
important to know all maximal subalgebras of Lie algebras of practical interest. Lie
algebras of semisimple type were classified a century ago in the work of Killing and
Cartan. Two important families of their maximal subalgebras were classified during 1950's
by Borel  and de Siebental \cite{BS}
(maximal rank reductive subalgebras) and by Dynkin (maximal
semisimple subalgebras) \cite{Dy}.

Quite different is the situation with solvable Lie algebras.
Classification of the bewildering variety of isomorphism classes
of such algebras remains today completely out of reach; only the
lowest dimensional ones ($\leq5$) have been described so far. A
curious sideline of our results illustrates the difficulty: In the
second part of this work, we describe over 80 non--isomorphic
solvable indecomposable Lie algebras of dimension 8. Compare that
with a single isomorphism class of semisimple Lie algebras of that
dimension!

Introduction of non-homomorphic relations between Lie algebras was motivated by the
practical need to relate meaningfully other pairs of Lie algebras. We are able to point out
only two very different relations of such kind, only the second one of the two is of
interest to us here. First is the subjoining of Lie algebras (see \cite{A,PSS,MPi} and
references therein). The second one is deformation of Lie algebras, introduced
independently in mathematics \cite{Seg} and in physics \cite{WI}. Intuitively
deformations can be defined as non-equivalent transformations of structure constants. For
rigid Lie algebras, like all the semisimple ones, that means singular transformations of
the constants. Subsequent development of the deformation theory in mathematics followed a
different path \cite{G} than in physics, showing so far only a marginal influence on the
latter.

In the physics literature there are several hundreds of papers dealing with
deformation/contraction of Lie algebras. Although their critical review would be a very
timely project, it is far beyond the scope of this article. Our immediate goal is to
apply recently invented variant of Wigner-In\"on\"u contractions, called graded
contractions \cite{PdeM}, to a specific case, the Lie algebra $sl(3,\C)$ equipped with
the Pauli grading \cite{PZ1}, and to find the outcome of all contractions preserving that
grading.

The set of $n\times n$ matrices, called here `generalized Pauli
matrices' \cite{PZ1}, has been known in mathematics long before
W. Pauli as a curious associative algebra,
or as a finite group of order $n^3$ \cite{D,S,weyl}.
In the physics literature the matrices are finding numerous
applications as well. Let us point out, for example,
\cite{GR,L,rausch,ST,V}.

During the last decade, based on the seminal paper \cite{PZ2},
important results were obtained in the classification of
gradings of classical simple Lie algebras \cite{HPP1,HPP2,HPP3}.
For a given simple Lie algebra it is thus possible ---  with
some effort --- to determine all its gradings. The extreme
gradings, which cannot be further refined (`fine gradings') are
useful because other gradings are obtained from them by suitable
combinations (`coarsening') of their grading subspaces. In
mathematics, the fine gradings of simple Lie algebras are
analogues of Cartan's root decomposition, which is one of them.
They define new bases with uncommon properties. In physics, they
provide maximal sets of quantum observables with additive quantum
numbers.

The general goal of our efforts, which goes well beyond the
present paper, is to find all the Lie algebras which can be
obtained from $sl(3,\C)$ by one--step contraction. That is,
excluding from consideration chains of successive contractions for
the same grading. There are four fine gradings of the algebra
\cite{HPPT}, i.e. gradings which cannot be further refined.
Therefore our task splits into four smaller ones, one for each
fine grading. All gradings of $sl(3,\C)$ are shown on Fig.~1,
including the four fine ones \cite{Pel}.

Best known of the four fine gradings \cite{HPPT} of $sl(3,\C)$ is
called the root decomposition or toroidal grading.
There the algebra is decomposed
as a linear space into the direct sum of eigenspaces of a maximal
torus of the group $SL(3,\C)$, or equivalently, into eigenspaces
of the corresponding Cartan subalgebra. All gradings, arising in
contractions preserving the toroidal grading, were found in
\cite{ALPW}.

The present case involves $sl(3,\C)$ decomposed into eigenspaces
of the adjoint action of generalized Pauli matrices. Although
the role of these matrices in grading $sl(n,\C)$ was singled out
only recently \cite{PZ1}, the special properties of the
associative algebra of these matrices were known long before
\cite{weyl} (also \cite{rausch} and references therein).
Remaining two gradings both involve outer automorphisms
of $sl(3,\C)$. Corresponding contractions have not been studied so far.

Solving the four contraction problems of $sl(3,\C)$ results in
four lists of contracted algebras. Subsequently the lists need
to be purged of overlaps. Relative practical difficulty of the
contraction problems can be read of Fig.~1. The graph of the
Figure shows successive refinements  of gradings leading from the
trivial one (whole $sl(3,\C)$) to the four fine gradings. Nodes of
the graph stand for non-equivalent gradings, links (arrows) indicate
refinements. The graph exhibits 8 levels corresponding to
numbers of grading subspaces: the numbers increase downwards, from 1 to 8, starting from the level of $sl(3,\C)$ itself.

In order to see the relative difficulty, we split
the graph into four overlapping subgraphs as follows. Retain in
each of the subgraphs only the links and nodes which connect one
fine grading with the whole $sl(3,\C)$. A node which appears on
several of the subgraphs indicates that the contracted Lie
algebras, corresponding to the grading of that node, appear on
several of the four lists. The less nodes a subgraph contains,
the more complicated  it is to find all contractions for that
fine grading. In such a comparison, description of the
contractions preserving the Pauli grading turns out to be a far
more challenging problem than the other three cases.

\begin{figure}
\setlength{\unitlength}{2pt}
\def\kr{\circle{7}}
\def\kb{\circle*{7}}
\begin{picture}(180,180)
\put(10,30){\kb} \put(50,30){\kb} \put(110,10){\kb}
\put(150,10){\kb} \put(130,30){\kr} \put(30,50){\kr}
\put(150,50){\kr} \put(50,70){\kr} \put(30,90){\kr}
\put(50,90){\kr} \put(90,90){\kr} \put(130,90){\kr}
\put(10,110){\kr} \put(110,110){\kr} \put(50,130){\kr}
\put(150,130){\kr} \put(110,150){\kr}
\put(110,159){\makebox(0,0){$sl(3,{\mathbb C})$}}
\put(10,22){\makebox(0,0){$\Gamma_a$}}
\put(50,22){\makebox(0,0){$\Gamma_b$}}
\put(110,2){\makebox(0,0){$\Gamma_c$}}
\put(150,2){\makebox(0,0){$\Gamma_d$}}
\put(170,159){\makebox(0,0){$M$}} \put(160,155){\line(1,0){20}}
\put(170,150){\makebox(0,0){$1$}}
\put(170,130){\makebox(0,0){$2$}}
\put(170,110){\makebox(0,0){$3$}} \put(170,90){\makebox(0,0){$4$}}
\put(170,70){\makebox(0,0){$5$}} \put(170,50){\makebox(0,0){$6$}}
\put(170,30){\makebox(0,0){$7$}} \put(170,10){\makebox(0,0){$8$}}
\put(106.679,148.893){\vector(-3,-1){53.359}}
\put(113.130,148.435){\vector(2,-1){33.74}}
\put(110,146.5){\vector(0,-1){33}}
\put(46.869,128.435){\vector(-2,-1){33.740}}
\put(48.435,126.869){\vector(-1,-2){16.870}}
\put(50,126.5){\vector(0,-1){33}} \put(52.475,
127.525){\vector(1,-1){35.050}}
\put(53.130,128.435){\vector(2,-1){73.740}}
\put(107.0878,108.0585){\vector(-3,-2){54.178}}
\put(110,106.5){\vector(0,-1){93}}
\put(111.941,107.0878){\vector(2,-3){36.12}}
\put(53.0012,68.199){\line(5,-3){93.999}} \put(145.284,
12.8297){\vector(2,-1){2}}
\put(148.435,126.870){\vector(-1,-2){16.86}}
\put(150,126.5){\vector(0,-1){73}}
\put(28.893,86.680){\vector(-1,-3){17.786}}
\put(30,86.5){\vector(0,-1){33}}
\put(11.107,106.680){\vector(1,-3){17.786}}
\put(48.435,86.870){\vector(-1,-2){16.870}}
\put(50,86.5){\vector(0,-1){13}}
\put(48.435,86.870){\vector(-1,-2){16.870}}
\put(52.8,87.9){\vector(4,-3){74.400}}
\put(91.941,87.088){\vector(2,-3){36.119}}
\put(130,86.5){\vector(0,-1){53}}
\put(32.47487372,47.52512628){\vector(1,-1){15.050}}
\put(50,66.5){\vector(0,-1){33}}
\put(132.475,27.525){\vector(1,-1){15.06}}
\put(150,46.5){\vector(0,-1){33}}
\end{picture}
\caption{The hierarchy of 17 gradings of $sl(3,\C)$
\cite{Pel}; $\Gamma_b$ and $\Gamma_c$ denote toroidal and
Pauli gradings, respectively. The gradings are distributed into 8 levels according to the number $M$
   of their grading subspaces. Nodes of
the graph stand for non-equivalent gradings, links (arrows) indicate refinements. Black circles denote fine gradings.}
\end{figure}
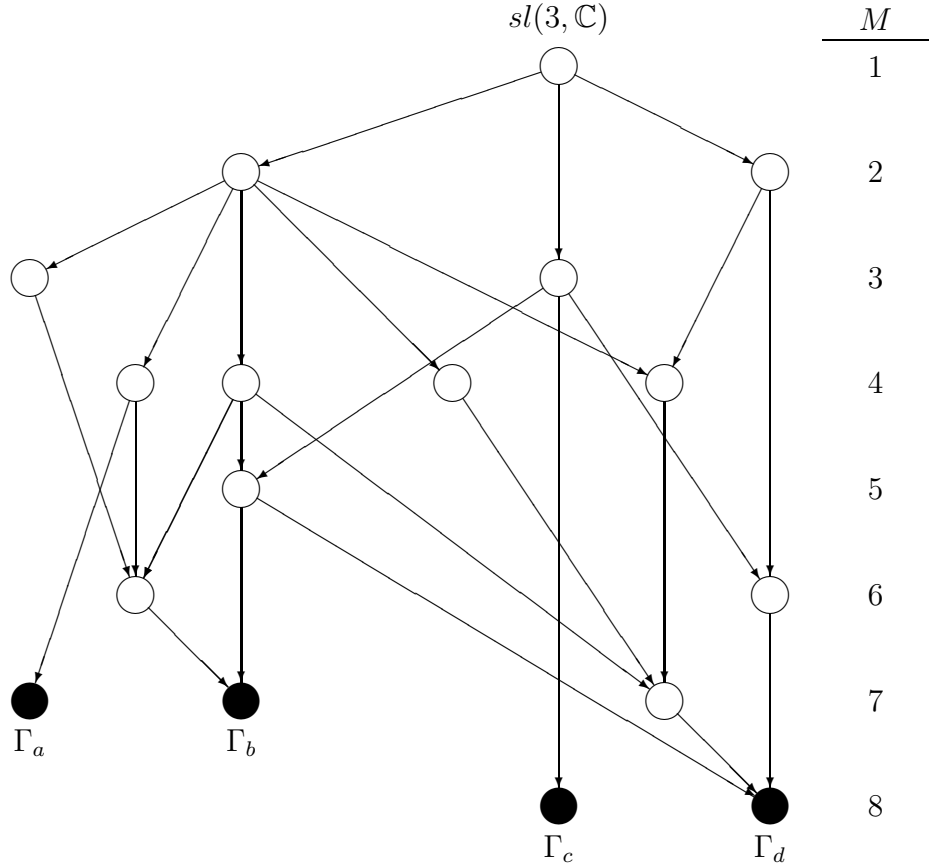

Specific aim of this article is to provide the list of 188 sets of
structure constants of the Lie algebras resulting from the
contractions of $sl(3,\C)$ which preserve the Pauli grading. On
this basis, isomorphism classes of contracted algebras are
determined.

A general new contribution of this paper in development of the
graded contraction method, is systematic exploitation of the
symmetries of the algebraic equations for the contraction
parameters while solving the equations.

The goal of this paper is to identify the Lie algebras from the
structure constants given by each contraction matrix. More
precisely, we determine the subset of those which are pairwise
non-isomorphic. The problem is particularly challenging for
solvable Lie algebras which are indecomposable and whose dimension
exceeds 5. Namely, little is known about these
 algebras, neither
their isomorphism classes, nor even estimates of their number.
Practically that means that existing algorithms for identification
of Lie algebras had to be extended to allow us to recognize
isomorphic pairs among Lie algebras arising from
 contractions. Here we shall restrict ourselves
to (finite-dimensional) Lie algebras over complex numbers $\bC$.

In the course of the work it turned out that finding an
isomorphism between algebras was a very non-trivial question. We
computed invariants for the algebras and when they were different,
the algebras were clearly non-isomorphic, but when they were the
same, new criteria had to be found. On the other hand we can try
to find an isomorphism in an explicit way. Corresponding system of
quadratic equations is generally more complicated than the system
of contraction equations and, moreover, the symmetries for this
system are not known. However, in some cases we were able to solve
this system on computer.

Our main method for identifying a Lie algebra given by its
structure constants was the paper by Rand, Winternitz and
Zassenhaus \cite{RWZ}.

Solutions of the system of contraction equations were written in
the form of $8\times 8$ contraction matrices $\varepsilon$ with 24
relevant entries. We start from the 8--dimensional Lie algebra
$sl(3,\bC)$ given by the structure constants $c_{i,j}^k$ in the
basis of the Pauli grading $\{e_i\}_{i=1}^8 $ :
\begin{equation}
[e_i,e_j] = \sum_{k=1}^8{c_{i,j}^ke_k}, \qquad i,j=1,\ldots,8.
\end{equation}
Then the solution $\varepsilon$ determines the contracted algebra
with the structure constants (in the same basis) given by
\begin{equation}
c(\varepsilon)_{i,j}^k = \varepsilon_{i,j}\ c_{i,j}^k.
\end{equation}

In Section 2, properties of gradings are briefly reviewed. The
Pauli grading of $sl(3,\C)$ is described explicitly, since it is
the starting point of our calculations.

In Section 3, we review the notion of a graded contraction as a
tool for obtaining new non-simple Pauli graded  8-dimensional Lie
algebras. Contrast between generic two-term contraction equations
and equations with three terms is underlined. A normalization
process and the concepts of continuous and discrete graded
contractions are introduced.

In Section 4, the notion of symmetry group of a grading is
introduced, its action on contraction matrices is defined, and it
is shown that the action induces the symmetries  of the system of
contraction equations. Crucial definition of equivalence of
solutions and the fact that Lie algebras obtained by graded
contractions corresponding to equivalent solutions are isomorphic,
is presented.

Section 5 contains the system of 48 two-term equations for the
Pauli grading of $sl(3,\C)$. We show how the symmetry group can
effectively help in reducing the number of equations.  Then all
higher-order identities \cite{WW} of order 2 and 3 are used
 to distinguish continuous contractions from the discrete ones.

In Section 6, Theorem \ref{main} is used to determine all
solutions of the contraction equations. A list of
nonequivalent cases is found in Appendix A.

In Section 7, seven steps of our algorithm used for identification
are described. In Section 8, the algorithm is applied to all
solutions of the contraction system. In order to identify the
contracted Lie algebras uniquely, the set of invariants was
supplemented in section 9 by dimensions of algebras of
derivations. The last section, Concluding remarks, contains a
number of comments, in particular there is an example of
three-term contraction equations and a number of interesting
problems is pointed out. In concluding remarks one finds a
comparison between the results of this paper and results for the
toroidal grading \cite{ALPW}. For completeness, the commutation
relations of all non-isomorphic contracted Lie algebras as well as
computed invariants are tabulated in Appendix B.

\section{Pauli grading of $sl(3,\C)$}

Pauli gradings of $sl(n,\C)$, for any $2\leq n<\infty$, were
described in \cite{PZ1}. In the case of $sl(3,\C)$ it is one of
the four fine gradings of that Lie algebra. The grading
decomposes  $sl(3,\C)$ into eight 1-dimensional subspaces.

In the defining 3-dimensional representation, basis
vectors/generators are $3\times 3$ generalized Pauli matrices:
\begin{align}\label{a2}
sl(3,\C) &=
L_{01}\oplus
L_{02}\oplus L_{10} \oplus L_{20}\oplus L_{11}\oplus L_{22} \oplus
L_{12}\oplus L_{21} \\ \label{zkos}
                     &= \C Q\oplus \C Q^2 \oplus \C P\oplus\C P^2
\oplus \C PQ \oplus \C P^2Q^2 \oplus \C PQ^2 \oplus \C P^2Q
\nonumber \\
                     &= \C \begin{pmatrix}
1 & 0 & 0 \\ 0 & \omega  & 0 \\ 0 & 0 & \omega ^{2}
\end{pmatrix}
\oplus\C
\begin{pmatrix}
1 & 0 & 0 \\ 0 & \omega ^{2} & 0 \\ 0 & 0 & \omega
\end{pmatrix}
\oplus\C
\begin{pmatrix}
0 & 1 & 0 \\ 0 & 0 & 1 \\ 1 & 0 & 0
\end{pmatrix}
 \oplus\C
\begin{pmatrix}
0 & 0 & 1 \\ 1 & 0 & 0 \\ 0 & 1 & 0
\end{pmatrix}
 \oplus \nonumber\\ & \oplus\C
\begin{pmatrix}
0 & \omega  & 0 \\ 0 & 0 & \omega ^{2} \\ 1 & 0 & 0
\end{pmatrix}
 \oplus\C
\begin{pmatrix}
0 & 0 & \omega  \\ 1 & 0 & 0 \\ 0 & \omega ^{2} & 0
\end{pmatrix}
 \oplus\C
\begin{pmatrix}
0 & \omega ^{2} & 0 \\ 0 & 0 & \omega  \\ 1 & 0 & 0
\end{pmatrix}
 \oplus\C
\begin{pmatrix}
0 & 0 & \omega ^{2} \\ 1 & 0 & 0 \\ 0 & \omega  & 0
\end{pmatrix} \nonumber
\end{align}
where $\omega = \exp (2 \pi \mathrm{i}/3)$.
Putting $L_{rs}:=\{X_{rs}\}_{lin}\equiv\C X_{rs}=\C Q^rP^s$, we
have the commutation relations,
\begin{equation}\label{komutator}
[X_{rs}, X_{r' s'}] =(\omega^{sr'}- \omega^{rs'})
      X_{r+r',s + s'\pmod3}\,.
\end{equation}
The index set $I$ for the Pauli grading consists of 8 couples $rs$, where $r,s=0,1,2$
with the exception of $00$.

Subsequently we make extensive use of the symmetry group
of the Pauli grading. It was described in detail in \cite{HPPT1} as
a finite matrix group
\begin{equation}\label{HH}
  H_3= \left\{\left( \begin{array}{cc}
               a&b\\
               c&d
              \end{array}
\right)\ \bigg|\  a,b,c,d\in\mathbb{Z}_3,\    ad-bc=\pm 1  \
\pmod 3 \right\}.
\end{equation}
It contains the subgroup of matrices with determinant +1 called
$SL(2,\Z_3)$. 

\section{Graded contractions of $sl(3,\C)$}

\subsection{Contraction parameters}\

The commutation relations \eqref{komutator} for basis elements
$X_{rs}$ are modified, for the purpose of a graded contraction
\cite{PdeM} of the Lie algebra, by introduction of the contraction
parameters $\varepsilon$,
\begin{align}\label{epskomutator}
[X_{rs}, X_{r' s'}]_\varepsilon: &=
\varepsilon_{(rs)(r's')}[X_{rs},X_{r' s'}] \\
&=\varepsilon_{(rs)(r's')}(\omega^{sr'}-\omega^{rs'})
      X_{r+r',s + s'\pmod3}\,.
\end{align}
Requirement, that the result of a contraction is a Lie
algebra, imposes certain conditions on the contraction parameters.
Antisymmetry of the modified commutator $[\,\,,\,\,]_{\ep}$
immediately gives
\begin{equation}\label{antisd}
  \varepsilon_{(rs)(r's')}=\varepsilon_{(r's')(rs)},
\end{equation}
hence it is convenient to view the set of contraction parameters
as a symmetric ($8\times 8$) {\bf contraction matrix}.
Among its 36 independent matrix elements, twelve are {\bf
irrelevant}, because in \eqref{epskomutator} they are multiplied by
vanishing commutators of $sl(3,\C)$. Consequently there are only
24 {\bf relevant} contraction parameters.
They are still subject to conditions imposed by the Jacobi
identities, i.e.
\begin{equation}\label{ceq}
e(i\:j\:k):\quad
[x,[y,z]_\ep]_\ep + [z,[x,y]_\ep]_\ep +[y,[z,x]_\ep]_\ep =0,
\end{equation}
where $x\in L_i$, $y\in L_j$, $z\in L_k$, $i=(rs)$, $j=(r's')$, $k=(r''s'')$. Each $
e(i\:j\:k)$ identifies a {\bf contraction equation}. We call the set of contraction
equations {\bf contraction system} $\es$, and the set of its solutions is denoted $\er (
\es )$. Using the Jacobi identity in $sl(3,\C)$,
\begin{equation}\label{eqpp2}
[x,[y,z]] + [z,[x,y]] +[y,[z,x]]=0
\end{equation}
one can rewrite (\ref{ceq}) in the form
\begin{align*}
e(i\:j\:k):\quad&(\ep_{i,j+k}\ep_{jk}-\ep_{k,i+j}\ep_{ij})[x,[y,z]]
+(\ep_{j,k+i}\ep_{ki}-\ep_{k,i+j}\ep_{ij})[y,[z,x]] =0 .
\end{align*}
In generic cases, when there exist
$x'\in L_i,\,y'\in L_j,\,z'\in L_k$ such that $[x',[y',z']]$
and $[y',[z',x']]$ are linearly independent, then equation
(\ref{ceq}) is equivalent to 2 two-term equations
\begin{equation}\label{eqpp3}
e(i\:j\:k):\quad
\ep_{i,j+k}\ep_{jk}=\ep_{k,i+j}\ep_{ij}=\ep_{j,k+i}\ep_{ki}.
\end{equation}
However, this condition is not always fulfilled. Then a three-term
contraction equations arises. See Concluding remarks for an
example.

\subsection{Equivalence transformations}\

The system $\es$ of quadratic equations admits many solutions. Our task is to find its
solutions which yield non-isomorphic Lie algebras  $\el^\ep$. It is a practical
imperative to make use of transformations which leave $\es$ unchanged, but transforming
otherwise the contraction parameters. In this subsection we introduce renormalization of
parameters $\ep$.

First we introduce, following \cite{PdeM}, {\bf commutative
elementwise matrix multiplication} denoted by $\bullet$. For two
matrices $A=(A_{ij}),\,B=(B_{ij})$ we define the matrix
$C:=(C_{ij})$ by the formula
\begin{equation}\label{componte}
  C_{ij}:=A_{ij}B_{ij} \q\q \mbox{(no summation)}
\end{equation}
and write $C=A\bullet B$.

For a given grading we renormalize grading subspaces, according to
$$
L_k\quad\longrightarrow\quad a_kL_k\,,\qquad
                   k\in I,\quad
                   0\neq a_k\in\C\,.
$$ Here $k$ takes value from an index set $I$ of the grading.
Matrix $\alpha:=(\alpha_{ij})$, where
\begin{equation}\label{alpe}
  \alpha_{ij}=\frac{a_i a_j}{a_{i+j}}\q \mbox{for}\,\, i,j\in I\,,
\end{equation}
is a {\bf normalization matrix}.

Normalization is a process based on the following lemma:
\begin{lem}\label{norma1}
Let $\el^\ep$ be a graded contraction of a graded Lie algebra $\el= \bigoplus _{i \in I}
L_i$. Then $\el^\mu$, where $\mu=\alpha\bullet\ep$, is for any normalization matrix
$\alpha$ a graded contraction of $\el$ and the Lie algebras $\el^\mu$ and $\el^\ep$ are
isomorphic.
\end{lem}

In many cases it is possible, by a suitable choice of the
constants $a_k$, $k\in I$, to transform matrix elements of $\ep$
to 1's and 0's only. We conclude in {\it Example} \ref{nonzer}
(Sec.~\ref{Finding}). that solution for the Pauli grading of
$sl(3,\C )$ which has all contraction parameters non-vanishing can
be always normalized to solution with 1's only. Clearly, this fact
is equivalent to:
\begin{prop}\label{normaun}
Every contraction matrix of the Pauli graded $sl(3,\C)$ without
zeros on relevant positions can be written in a form of
normalization matrix (\ref{alpe}).
\end{prop}

\subsection{Continuous and discrete graded contractions}\

There are solutions of two types of a given contraction system
$\es$, continuous and discrete ones.

 A solution $\ep\in \er(\es )$ is {\bf continuous} if there exists a
continuous set of solutions $\ep(t)\in \er(\es),\, 0<t\leq1$, such that, for all relevant
contraction parameters, one has
\begin{equation}\label{spoj}
\ep_{ij}(1)=1,\qquad
\ep_{ij}(t)\neq0,\qquad
\ep_{ij}= \lim_{t\rightarrow 0}\ep_{ij}(t).
\end{equation}
If a solution is not continuous, then it is called {\bf discrete}.

If, for a given continuous graded contraction
$\ep\in \er(\es )$, the corresponding continuous set
of solutions has the form
\begin{equation}\label{spoj2}
\ep_{ij}(t)=t ^{n_i+n_j-n_{i+j}} \q i,j\in I,\q n_i,n_j,n_{i+j}
\in \Z,
\end{equation}
then $\ep$ is a generalized In\"{o}n\"{u}-Wigner contraction
\cite{WW}.

Traditionally continuous contractions were more thoroughly
studied in the literature \cite{WW}.
Let us point out higher-order identities which we use
in Sect. \ref{higher} as an effective tool
for identifying {\it discrete} graded contractions in
the present case.

\section{Symmetries and graded contractions}
The contraction system $\es$ for the Pauli grading of $sl(3,\C)$
is the system of 48 quadratic equations in 24 variables. Making
use of the symmetries should simplify the solution of such a large
system.

Symmetries are involved in three different ways in our problem. First it is the symmetry
group of the Pauli grading of $sl(3,\C)$. Second is the symmetry of the system $\es$ of
contraction equations. And third, the symmetries transforming solutions among themselves.


\subsection{Symmetry group of the Pauli grading}\

For general gradings, symmetry groups were introduced in
\cite{PZ2} and, for Pauli gradings of $sl(n,\C)$ were described in
\cite{HPPT1}. We recall in particular the following:

The symmetry group $\aut\Gamma$ of a grading $\Gamma : \el=\bigoplus _{i \in I} L_i$ is
defined as a subgroup of $\aut \el$ which contains automorphisms $g$ with the property $g
L_i=L_{\pi_g(i)},$ where $\pi_g :I \rightarrow I$ is a permutation of the index set $I$.

This group is described in detail in \cite{HPPT1,HPPT2,HPPT}, where
an important theorem was proved; we specify it for $n=3$:
\begin{thm}\label{quo}
{\it
The symmetry group of the Pauli grading of $sl(3,\C)$ is isomorphic
to the matrix group $H_3$. It is a finite matrix group
\begin{equation}\label{HH3}
  H_3= \left\{A=\left( \begin{array}{cc}
               a&b\\
               c&d
              \end{array}
\right)\ \bigg|\  a,b,c,d\in\mathbb{Z}_3,\ ad-bc=\pm 1\pmod3
\right \}.
\end{equation}

Let $\pi_A$ denote the permutation corresponding to a matrix $A\in
H_3$; then the action of $\pi_A$ on the grading indices is given by
\begin{equation}\label{per5}
  \pi_A\,(i\,\,j)=(i\,\,j)\left( \begin{array}{cc}
               a&b\\
               c&d
              \end{array}
\right),
\end{equation}
where matrix multiplication modulo $3$ is understood.}
\end{thm}
Note that $H_3$ contains the subgroup $SL(2,\Z_3)$ of matrices with
determinant +1.

\subsection{Action of the symmetry group of the grading}\

Let us define a set of {\bf relevant pairs of grading indices}
$\I$ by
\begin{equation}\label{ungrad}
\I := \left\{ i\:j\,\big|\,i,j\in I,\,[L_i,L_j]\neq \{0\} \right\}.
\end{equation}
For the Pauli grading of $sl(3,\C)$ we obtain explicitly
\begin{equation}\label{IN}
\I=\left\{\,(ij)(kl)\,\big|\,\,jk-il\neq 0\,(\mbox{mod
}3),\,\,(ij),(kl)\in\Z_3\times\Z_3\backslash \{(0,0)\}\right\}
\end{equation}
by analyzing relations \eqref{komutator}. A set of {\bf relevant
contraction parameters} $\ep_{ij}$, due to (\ref{antisd}), can be
written as $\E:=\{ \ep_k,\,k \in \I\}$. For a permutation $\pi$
and a contraction matrix $\ep=(\ep_{ij})$, an {\bf action of $\pi$
on a contraction matrix} $\ep \mapsto \ep^\pi$ is defined by
\begin{equation}\label{action}
 (\ep^\pi)_{ij}:=\ep_{\pi(i)\pi(j)}.
\end{equation}
We observe that an {\bf action on variables}
$\ep_{ij}\mapsto\ep_{\pi(i)\pi(j)}$ is in fact the action on the
set of relevant variables $\E$: if
$\ep_{ij}\in\E,\,[L_i,L_j]\neq\{0\}$ and $g\in\aut\Gamma$, $\pi$
the corresponding permutation, then $\{0\}\neq
g[L_i,L_j]=[L_{\pi(i)},L_{\pi(j)}]$ and $\ep_{\pi(i)\pi(j)}\in\E$.
Hence even if the matrix $\ep^\pi$ has, in general, zeros on
different positions than matrix $\ep$, the matrix $\ep^\pi$ has
zeros on the same irrelevant positions as the matrix $\ep$, i.e.
the irrelevant positions are fixed.

We verify that \eqref{action} is a well-defined action of the
group $\aut\Gamma$ on the set $\er(\es)$ of solutions of the
contraction system. However, in Subsect. \ref{ceqi} we are going
to widen the definition of equivalence on the solutions. The
following lemma shows that the action of $\pi$ induces an
isomorphism of Lie algebras associated with contraction matrices
$\ep$ and $\ep^\pi$.

\begin{lem}\label{norma2}
{\it Let $\el^\ep$ be a graded contraction of a $\Gamma$--graded Lie algebra
$\el=\bigoplus _{i \in I} L_i$. Then $\el^{\ep^\pi}$ is, for permutations $\pi$
corresponding to elements of $\aut\Gamma$, also a graded contraction of $\el$ and the Lie
algebras $\el^{\ep^\pi}$ and $\el^\ep$ are isomorphic, $\el^{\ep^\pi}\simeq \el^\ep$.}
\end{lem}
\begin{pf}
For  $g\in\aut\Gamma$ take the corresponding $\pi$. Consider
\begin{equation}\label{permg}
gx = z,\ gy= w,\q \,x\in L_i,\,y\in L_j,\,z\in L_{\pi (i)},\,w\in L_{\pi(j)},\,i,j\in I.
\end{equation}
Then for all $x$, $y$ the bilinear mapping
${[x,y]_{\ep^\pi}=\ep_{\pi(i)
\pi (j)} [x,y]}$ and the Lie bracket ${[x,y]_{\ep}=\ep_{ij} [x,y]}$
satisfy
\begin{equation}\label{permiz}
[x,y]_{\ep^\pi}=\ep_{\pi (i) \pi (j)} [x,y]=\ep_{\pi (i) \pi
(j)}g^{-1}[z,w]=g^{-1}[g x,g
y]_{\ep}.
\end{equation}
Hence $\el^{\ep^\pi}$ is a Lie algebra and $g$ is an isomorphism
between $\el^{\ep^\pi}$ and $\el^\ep$.
\end{pf}

\subsection{Symmetries of the system of contraction equations}\

It follows from Lemma \ref{norma2} that for a given solution $\ep$
one can construct new contraction matrices $\ep^\pi$ which are
again solutions of the contraction system (and correspond to
isomorphic Lie algebras). We thus obtained the substitutions
$\ep_{ij}\mapsto\ep_{\pi(i)\pi(j)}$ preserving the set
of solutions of the contraction system.

Now we can define an {\bf action of $\aut\Gamma$ on the
contraction system $\es$}: each equation in $\es$ is labeled by a
triple of grading indices and we write $e(i\:j\: k)\in \es$ in the
form
\begin{equation}\label{sys1}
e(i\:j\:k):\, [x,[y,z]_\ep]_\ep + \mbox{cyclically} =0,\q  x\in L_i, y\in L_j,z \in L_k;
\end{equation}
then for each $\pi$ we define the action
\begin{equation}\label{akc}
e(i\: j\: k)\,\mapsto\, e(\pi(i)\: \pi(j)\: \pi(k)).
\end{equation}
Note that equation $e(\pi(i)\: \pi(j)\: \pi(k))$ can be written
for each $g$ and the corresponding $\pi$ as
\begin{equation}\label{jac1}
e(\pi(i)\:\pi(j)\:\pi(k)):\,
[gx,[gy,gz]_\ep]_\ep + \mbox{cyclically}=0.
\end{equation}
 According to (\ref{permiz}) this is equal to
\begin{equation}\label{jac2}
g[x,[y,z]_{\ep^\pi}]_{\ep^\pi} + \mbox{cyclically} =0
\end{equation}
and (\ref{jac2}) is satisfied if and only if
\begin{equation}\label{jac3}
[x,[y,z]_{\ep^\pi}]_{\ep^\pi} + \mbox{cyclically} =0.
\end{equation}

Equation (\ref{jac3}) is precisely the equation (\ref{sys1})
after the substitution $\ep_{ij}\mapsto\ep_{\pi(i)\pi(j)}$. In
this way we have not only verified the invariance of the
contraction system (up to equivalence of equations), but also have
shown the method which is subsequently used for its construction.
Namely, having chosen a starting equation one can write a whole
orbit of equations merely by substituting
$\ep_{ij}\mapsto\ep_{\pi(i)\pi(j)}$ till all permutations
$\pi$ corresponding to $\aut\Gamma$ are exhausted.

If we denote unordered k-tuple of grading indices $i_1,i_2,\dots,i_k\in I$ as $i_1\: i_2
\dots i_k$ and for permutation $\pi$ corresponding to $g \in\aut\Gamma$ define an {\bf
action on unordered k-tuples} as
\begin{equation}\label{akc0}
i_1 \:i_2 \:\dots i_k\,\mapsto\, \pi(i_1)\: \pi(i_2) \dots \pi(i_k),
\end{equation}
then it is clear that orbits of equations correspond to orbits of
unordered triples of grading indices. In the following, we shall
write also $\pi \in \aut \Gamma$ for a permutation corresponding
to some $g\in \aut \Gamma$.

\subsection{Equivalence of solutions}\label{ceqi}\

Combining Lemma \ref{norma1} and Lemma \ref{norma2} it is easy to
see that an equivalence relation on the set $\er (\es)$ naturally
arises: we define two solutions $\ep',\ep \in \er(\es)$ to be {\bf
equivalent}, $\ep'\sim\ep$, if there exists a normalization matrix
$\alpha$ and a permutation $\pi$ (corresponding to
an element of $\aut\Gamma$) such that
\begin{equation}\label{ekkk}
\ep'=\alpha\bullet\ep^\pi.
\end{equation}
Note that this definition of equivalence is different from
that of \cite{WW}, Definition 2.2. For instance it is clear
that two equivalent solutions {\it do not} have the same set
of zeros. It follows from Lemma \ref{norma1} and Lemma
\ref{norma2} that the following Proposition holds:
\begin{prop}\label{cor1}
 {\it Lie algebras corresponding to equivalent
solutions of the system of contraction equations are isomorphic.}
\end{prop}

\section{Contraction system for the Pauli grading of
$sl(3,\C)$}\label{systemcon}
In this section we describe the system $\es$ of 48 quadratic
contraction equations for the Pauli grading of $sl(3,\C)$ using the
symmetry group.

 Let us take the Pauli
grading in the form \eqref{a2}, the grading group is $\Z_3
\times \Z_3$; no subspace is labeled by $(0,0).$ We choose the
explicit form of matrix $\ep$ defined by \eqref{epskomutator}; it
is an $8\times 8$ symmetric matrix with $24$ relevant variables
\begin{equation}\label{epsi}
\ep=
 \begin{pmatrix}
   0 & 0 & \ep_{(01)(10)} & \ep_{(01)(20)} & \ep_{(01)(11)} &
\ep_{(01)(22)} & \ep_{(01)(12)} & \ep_{(01)(21)} \\
   0 & 0 & \ep_{(02)(10)} & \ep_{(02)(20)} & \ep_{(02)(11)} &
\ep_{(02)(22)} & \ep_{(02)(12)} & \ep_{(02)(21)} \\
   \ep_{(01)(10)} & \ep_{(02)(10)} & 0 & 0 & \ep_{(10)(11)} &
\ep_{(10)(22)} & \ep_{(10)(12)} & \ep_{(10)(21)} \\
   \ep_{(01)(20)} & \ep_{(02)(20)} & 0 & 0 & \ep_{(20)(11)} &
\ep_{(20)(22)} & \ep_{(20)(12)} & \ep_{(20)(21)} \\
   \ep_{(01)(11)} & \ep_{(02)(11)} & \ep_{(10)(11)} & \ep_{(20)(11)} & 0 &
0 & \ep_{(11)(12)} & \ep_{(11)(21)} \\
   \ep_{(01)(22)} & \ep_{(02)(22)} & \ep_{(10)(22)} & \ep_{(20)(22)} & 0 &
0 & \ep_{(22)(12)} & \ep_{(22)(21)} \\
   \ep_{(01)(12)} & \ep_{(02)(12)} & \ep_{(10)(12)} & \ep_{(20)(12)} &
\ep_{(11)(12)} & \ep_{(22)(12)} & 0 & 0 \\
   \ep_{(01)(21)} & \ep_{(02)(21)} & \ep_{(10)(21)} & \ep_{(20)(21)} &
\ep_{(11)(21)} & \ep_{(22)(21)} & 0 & 0
 \end{pmatrix}.
\end{equation}

Contraction equations $e( (i j)\: (k l)\: (m n))\in \es$ should hold for all possible
triples of indices $(i j),(k l),(m n)$. It is clear that for each triple, whose two
indices are identical, the equation is identically fulfilled. Equations also do not
depend on the ordering of the triples. The number of equations is then equal to the
combination number $\left(\begin{smallmatrix} 8 \\ 3 \end{smallmatrix}\right)=56$.
Moreover, equations with $i+k+m=0$ and $j+l+n=0$ simultaneously are also fulfilled
identically. This situation arises in eight cases. Hence the contraction system consists
of $48$ equations.

Let us show an example of an equation computed for a chosen
triple, say $(01)(02)(10)$. The result is
\begin{equation}\label{pr}
[\ep_{(02)(10)}\ep_{(01)(12)}(\omega^2-1)(\omega
-1)+0+\ep_{(10)(01)}\ep_{(02)(11)}(1-\omega
)(\omega^2-1)]X_{10}=0.
\end{equation}

Recall that the symmetry group $H_3$ has 48 elements. It turns out
that its application on the triples leads to exactly {\it two}
orbits of its action, each consisting of 24 distinct triplets.
We choose the triples {(01)(02)(10)} and {(01)(10)(11)} as
representative elements of the two orbits. We observe that all 24
elements of each orbit are obtained by the action of 24
elements of the subgroup $SL(2,\Z_3)\subset H_3$ starting from an
arbitrary point. Then for our choice of representative points, our
system $\es$ can be written simply as
\begin{align}
\es^a\,:\,\,&\ep_{(02)(10)A}\ep_{(01)(12)A}-\ep_{(01)(10)A}\ep_{(02)(11)A}=0&\label{fff}
\\
\es^b\,:\,\,
&\ep_{(10)(11)A}\ep_{(01)(21)A}-\ep_{(01)(11)A}\ep_{(10)(12)A}=0&\q\forall
A \in SL(2,\Z_3)\label{ffff}
\end{align}
where we used the abbreviation
$\ep_{(i j)(k l)A}:=\ep_{(ij)A,(kl)A}$.

It turns out that the system $\es^a$ contains dependent equations
which can be eliminated. This is seen  as follows. Equation
obtained from \eqref{fff} by the 'action' of the unit matrix
can also be written in the form
\begin{equation}\label{X1}
\ep_{(01)(10)X}\ep_{(02)(11)X}-\ep_{(01)(10)}\ep_{(02)(11)}=0,\q\q
X=\begin{pmatrix}
  1 & 2 \\
  0 & 1
\end{pmatrix}
\end{equation}
This is due to the fact that the quadruples of indices
$[(02)(10)][(01)(12)]$ and $[(01)(10)][(02)(11)]$ lie in the same
$SL(2,\Z_3)$--orbit (the pairs of indices in brackets $[\,,\,]$ and
the pairs of these brackets are unordered). The equation generated
from equation \eqref{X1} by the matrix $A=X$
\begin{equation}\label{X2}
\ep_{(01)(10)X^2}\ep_{(02)(11)X^2}-\ep_{(01)(10)X}\ep_{(02)(11)X}=0
\end{equation}
is also contained in $\es^a$, due to \eqref{fff}. By adding
equations \eqref{X1} and \eqref{X2} we get
\begin{equation}\label{X3}
\ep_{(01)(10)X^2}\ep_{(02)(11)X^2}-\ep_{(01)(10)}\ep_{(02)(11)}=0.
\end{equation}
Since $X^3=1$ holds, equation \eqref{X3} is generated from
equation \eqref{fff} by matrix $A=X^2$. Hence we conclude
that the left cosets of $SL(2,\Z_3)$ with respect to the
cyclic subgroup $\{1,X,X^2\}$ generate the triples of
dependent equations. By Lagrange's theorem, the number of these
cosets is $24/3=8$. In this way we obtained 8 equations
(one to each coset) which can be eliminated from the system
$\es^a$. Concerning $\es^b$, we observe that the
quadruples of indices $[(10)(11)][(01)(21)]$ and
$[(01)(11)][(10)(12)]$ {\it do not} lie in the same orbit.
Therefore the equations of $\es^b$ are independent.

\subsection{Higher--order identities}\label{higher}
Higher--order identities will be useful for distinguishing between
continuous and discrete contractions. A {\bf higher--order
identity of order $k$} is defined \cite{WW} as an equation of
the type
\begin{equation}\label{hihf}
\ep_{i_1}\ep_{i_2}\cdots\ep_{i_k}=\ep_{j_1}\ep_{j_2}\cdots\ep_{j_k};
\end{equation}
here $k\in\N$, ${i_1},{i_2}\cdots{i_k}$ and
${j_1},{j_2}\cdots{j_k}$ are disjoint sets of relevant pairs of
grading indices, and equation \eqref{hihf} holds for all
contraction matrices $\ep$ without zeros on all relevant positions
(i.e., $\ep_i\neq 0$ for all $i \in \I$), but is violated by some
contraction matrix with zero on some relevant position. It is
obvious that continuous solutions, as limits of non-zero
solutions, do have to satisfy all higher--order identities (see
also Theorem 5.1 \cite{WW}):
\begin{prop}\label{coccc}
Let a solution $\ep\in\er (\es)$ of the contraction system $\es$
be a continuous graded contraction. Then $\ep$ satisfies all
higher--order identities.
\end{prop}
It follows that those solutions which violate any higher--order
identity are discrete.
\subsection{Higher--order identities for the Pauli grading
 of $sl(3,\C)$} \

As an example of a third order identity for the Pauli
grading one can give the following equation:
\begin{equation}\label{hodd}
\ep_{(01)(10)} \ep_{(01)(12)} \ep_{(02)(21)} =
\ep_{(01)(22)}\ep_{(01)(21)} \ep_{(02)(12)}.
\end{equation}
It can either be deduced directly from the system $\es$,
or we can use the fact (Proposition \ref{normaun})
that all solutions with all non-zero relevant elements can
be written in the form \eqref{alpe} of the normalization
matrix $\alpha$. Hence the identity
\begin{equation}\label{apaa}
\frac{a_{(01)}a_{(10)}}{a_{(11)}}\frac{a_{(01)}a_{(12)}}{a_{(10)}}\frac{a_{(02)}a_{(21)}}{a_{(20)}}=
\frac{a_{(01)}a_{(22)}}{a_{(20)}}\frac{a_{(01)}a_{(21)}}{a_{(22)}}\frac{a_{(02)}a_{(12)}}{a_{(11)}}
\end{equation}
is evidently satisfied. We observe that the identity \eqref{hodd}
is evidently violated for instance by the contraction matrix
equivalent to $\ep_{21,4}$ given in Appendix:  $\ep_{(01)(10)}=\ep_{(01)(12)}= \ep_{(02)(21)} =1$
and the other $\ep$'s are equal to zero.

Applying the symmetry group $H_3$ to \eqref{hodd},
we can write the 24-point orbit of higher-order identities
in the form
\begin{equation}\label{hodd2} \ep_{(01)(10)A} \ep_{(01)(12)A}
\ep_{(02)(21)A} = \ep_{(01)(22)A}\ep_{(01)(21)A} \ep_{(02)(12)A}
\q\q \forall A\in H_3
\end{equation}
Note that the action is effective only for 24 elements
of the subgroup $SL(2,\Z_3)$.

We have found a set of second and third order identities.
In all we have 104 such identities, 24 of them being
of second order. Table 1 lists their representative points
 and the numbers of the resulting identities under the action of $SL(2,\Z_3)$.

For each solution of the system $\es$ we were able
to decide that one of two exclusive alternatives holds:
\begin{itemize}
\item
 either we found that a solution violates some of 104
 identities listed in Table 1 and therefore it is
 {\it discrete}
\item
 or we explicitly found a continuous path of the form
 (\ref{spoj2}), hence the solution is {\it continuous}
\end{itemize}
Thus is was not necessary to investigate the completeness
of our set of higher-order identities.
\begin{table}[h!]
\small
\begin{tabular}[b]{|c||c||c|}
\hline
 \parbox[l][20pt][c]{5pt}{}Order & Representative equation&Number of
equations \\ \hline\hline
\parbox[l][23pt][c]{1pt}{} 2& $ \ep_{(01)(10)} \ep_{(02)(11)}
=\ep_{(01)(20)}\ep_{(02)(21)}$ & \it{24}\\ \hline
\parbox[l][23pt][c]{1pt}{} 3& $ \ep_{(01)(10)} \ep_{(01)(11)}
\ep_{(01)(12)} =\ep_{(01)(20)}\ep_{(01)(22)} \ep_{(01)(21)}$ &
\it{8}\\ \hline
  \parbox[l][23pt][c]{1pt}{} 3& $ \ep_{(01)(10)} \ep_{(01)(12)}
\ep_{(02)(21)} =\ep_{(01)(22)}\ep_{(01)(21)} \ep_{(02)(12)}$&
\it{24} \\ \hline
  \parbox[l][23pt][c]{5pt}{} 3& $ \ep_{(01)(10)} \ep_{(01)(11)}
\ep_{(02)(21)} =\ep_{(01)(22)}\ep_{(01)(21)} \ep_{(02)(10)}$ &
\it{24}\\ \hline
  \parbox[l][23pt][c]{5pt}{} 3& $ \ep_{(01)(10)} \ep_{(01)(11)}
\ep_{(02)(22)} =\ep_{(01)(20)}\ep_{(01)(22)} \ep_{(02)(10)}$  &
\it{24} \\ \hline
\end{tabular}
\medskip
\centering \caption[l]{ {\it Orbits of 2nd and 3rd order
identities for the Pauli grading} } \vspace{-8pt}\label{tab}
\end{table}

\section{Solution of the system $\es$ of contraction
equations for the Pauli grading of $sl(3,\C)$}

The goal of this section is to determine all equivalence classes
(in the sense of Sect. \ref{ceqi}) of solutions of the nonlinear
contraction system. Since the method of solution proposed by
\cite{Spain} was inapplicable in our case, we had to find another
way. Of course, a laborious case by case analysis as in
\cite{ALPW} is always possible. However, we describe a method
simplifying this laborious analysis. Our approach takes advantage
of the symmetries of the Pauli grading, more precisely, the
symmetries of the system of contraction equations as well as the
symmetries defining equivalence classes of solutions. We obtained
the complete set of (equivalence classes of) solutions, among them
13 which depend on one or two parameters.

\subsection{The algorithm for solving the system $\es$} \

Our method is based on the fact that under suitable assumptions, the
system $\es$ can be easily explicitly solved. But after leaving the
assumption, i.e. putting zero on the position we had assumed
non--zero before, the solving becomes far more complicated. We shall
formulate an algorithm allowing us to bypass this problem. It is
based on the following theorem employing our notion of equivalent
solutions \eqref{ekkk}.
\begin{thm}\label{main}
Let $\er (\es)$ be the set of solutions and $\I$ the set of
relevant pairs of unordered indices of the contraction system
$\es$ of a graded Lie algebra $\el=\bigoplus _{i\in I}L_i$. For
any subsets $\qe \subset \er (\es)$ and
$\pe=\{k_1,\,k_2,\dots,k_m\}\subset\I$ we denote
\begin{align*}
\er_0&:=\left\{\ep\in\er(\es)\big|(\forall
\ep'\in\qe)(\ep\nsim\ep') \right\}\\ \er_1&:=\left\{ \ep \in
\er_0\, \big|\,(\forall k \in \pe )( \ep_k\neq 0 )\right\}.
\end{align*}
Then a solution $\ep\in\er_0$ is not equivalent to any solution
in $\er_1$ {\it if and only if} the following system of equations
holds:
\begin{align}\label{none}
\ep_{\pi_1 (k_1)} \ep_{\pi_1 (k_2)}\cdots \ep_{\pi_1 (k_m)}&=0
\nonumber\\ \vdots \q\q&
\\ \ep_{\pi_n (k_1)} \ep_{\pi_n (k_2)}\cdots \ep_{\pi_n (k_m)}&=0;
\nonumber
\end{align}
here the set of permutations $\{\pi_1,\pi_2,\dots,\pi_n\}$ exhausts
all elements of the symmetry group $\aut\Gamma$ of the grading.
\end{thm}
\begin{pf}
For any $\ep\in\er_0$ we have (see \ref{ekkk}):
\begin{align}
(\exists\ep'\in\er_1)(\ep\sim\ep')&\Leftrightarrow \,\,(\exists
\ep'\in\er_1)(\exists\alpha)(\exists\pi\in\
\aut\Gamma)(\alpha\bullet\ep^\pi=\ep')\label{main1}\\
&\Leftrightarrow\,\,(\exists\alpha)(\exists\pi\in\ \aut\Gamma)(\alpha\bullet\ep^\pi \in
\er_0\  \text{and}\ (\alpha\bullet\ep^\pi)_k\neq 0,\,\, \forall k \in\pe) \label{main2}\\
&\Leftrightarrow\,\,(\exists\pi\in\ \aut\Gamma)(\forall k \in \pe)((\ep^\pi)_k\neq
0).\label{main3}
\end{align}
The equivalence \eqref{main1} directly follows from the
definition \eqref{ekkk}, the equivalence \eqref{main2} expresses
the trivial fact that
$(\exists\ep'\in\er_1)(\alpha\bullet\ep^\pi=\ep')\Leftrightarrow
(\alpha\bullet\ep^\pi \in \er_1)$. Since $\alpha\bullet\ep^\pi \in
\er_0$ is for any $\ep\in\er_0$ automatically fulfilled and
$(\alpha\bullet\ep^\pi)_k\neq 0\Leftrightarrow(\ep^\pi)_k\neq 0$,
the equivalence (\ref{main3}) follows.

Negating \eqref{main3} we obtain
\begin{align*}
(\forall\ep'\in\er_1)(\ep\nsim \ep')\Leftrightarrow
\,\,(\forall\pi\in\ \aut\Gamma)(\exists k \in
\pe)((\ep^\pi)_k = 0)
\end{align*}
and this is the statement of the theorem.
\end{pf}
We call the system of equations \eqref{none} corresponding to the
sets $\qe \subset \er (\es)$ and $\pe\subset\I$ a~{\bf
non-equivalence system}.

Repeated use of the theorem leads to the following algorithm
for the evaluation of solutions:
\begin{enumerate}

\item we set $\qe=\emptyset$ and suppose we have a set of
assumptions $\pe^0\subset\I$. Then
$\er_0=\er(\es)$, and we explicitly evaluate
$$\er^0=\left\{
\ep\in \er(\es)\,\,|\,\,(\forall k \in \pe^0)(\ep_k \neq 0)
\right\} $$
and write the {\it non-equivalence system} $\es^0$ of
equations \eqref{none} corresponding to  $\qe=\emptyset,\,\pe^0$.

\item we set $\qe=\er^0$ and suppose we have the set
$\pe^1\subset\I$. Then
$\er_0=\er(\es\cup\es^0)$; furthermore we explicitly evaluate
$$\er^1=\left\{ \ep\in \er(\es\cup\es^0)\,\,|\,\,(\forall k \in
\pe^1)(\ep_k \neq 0) \right\} $$ and write the non-equivalence
system $\es^1$ corresponding to $\qe=\er^0,\,\pe^1$.

\item we set $\qe=\er^0\cup\er^1$. Then
$\er_0=\er(\es\cup\es^0\cup\es^1)$ and we continue till we have
evaluated the whole $\er(\es)$ up to equivalence, i.e. till we have
arrived at such $\qe$ that the corresponding set $\er_0$ is empty or
trivial. Thus the idea is to repeatedly enlarge the set $\qe$ for as
long as possible.
\end{enumerate}

It is clear that the algorithm crucially depends on the choice of
the subsets $\pe^0,\,\pe^1,\dots$ of $\I$. Since the system
$\es$ can be solved explicitly under the assumption that two of its
variables do not vanish, we develop a theory for pairs from $\I$.
For fixed $k \in \I$ we define an equivalence relation $\equiv^k$ on
the set $\I^k:=\I\setminus \{k\}$: \\
for $i,j\in\I^k$
\begin{equation}\label{ekviv}
i \equiv^k j\,\,\,\Leftrightarrow\,\,
(\exists\pi\in\aut\Gamma)
(\pi\left(i\,\,k\right)=\left( j\,\,k\right) ),
\end{equation}
where $(i\,\,j)$ denotes an (unordered) pair $i,j\in\I$ and
$\pi\left( i\,\,k\right):=\left(\pi(i)\,\,\pi(k)\right)$.

The application of this equivalence will be seen in our concrete
evaluation in the next subsection. We will make use of the
following fact: the set of relevant indices $\I$ has $24$ elements
which are explicitly written in the matrix \eqref{epsi}. We choose
the index $k=(01)(10)$ and in Table 2 we list nine equivalence
classes $\I_1^k,\dots,\I_9^k$ of the equivalence $\equiv^k$:

\begin{table}[h!]
\small
\begin{tabular}[b]{|c||l|}\hline
\parbox[l][20pt][c]{3pt}{}$ \I_1^k$ & (11)(12), (11)(21),
(22)(12), (22)(21) \\ \hline
\parbox[l][20pt][c]{3pt}{}$ \I_2^k$ & (01)(11), (10)(11), (01)(12),
(10)(21) \\ \hline \parbox[l][20pt][c]{3pt}{}$\I_3^k$ & (02)(22),
(20)(22), (02)(21), (20)(12) \\ \hline
\parbox[l][20pt][c]{3pt}{}$\I_4^k$ & (01)(20), (02)(10) \\ \hline
\parbox[l][20pt][c]{3pt}{}$\I_5^k$ & (01)(22), (10)(22) \\ \hline
\parbox[l][20pt][c]{3pt}{}$\I_6^k$ & (01)(21),
(10)(12)
\\ \hline \parbox[l][20pt][c]{3pt}{}$\I_7^k$ & (02)(11), (20)(11) \\
\hline \parbox[l][20pt][c]{3pt}{}$\I_8^k$ & (02)(12), (20)(21)
\\ \hline
\parbox[l][20pt][c]{3pt}{}$\I_9^k$ & (02)(20) \\ \hline
\end{tabular}
\medskip
\centering \caption[l]{{\it The equivalence classes of
$\equiv^{(01)(10)}$}} \vspace{-8pt}\label{tab0}
\end{table}

\subsection{Finding solutions of $\es$} \label{Finding}\

The solutions are found in five consecutive steps. In each of
the following steps, $k=(01)(10)$ is fixed, and it is assumed that
the corresponding $\ep_k\neq 0$. Let in $\er^m$ further $\ep_i \neq
0$ be assumed. Then at the next step, in evaluating $\er^{m+1}$, one
finds the following: the non--equivalence system $\es^m$ and previous
assumption $\ep_k\neq 0$ imply zeros on all positions $j,\,j\equiv^k
i$. It is advantageous to take pairs of indices from the three
largest classes in Table 2, namely,
$(22)(21)\in\I_1^k$, $(10)(11)\in\I_2^k$, $(02)(22)\in
\I_3^k$, in order to evaluate  the sets $\er^0,\,\er^1$ and
$\er^2,\,\er^3$ respectively.

We first list the five steps and then
make more detailed explanations.
\begin{align*}
  1.\q\q &\er^0=\left\{ \ep\in \er(\es)\,\,|\,\,\ep_{(01)(10)}\neq
0,\,\ep_{(22)(21)}\neq 0 \right\}\\
&\es^0:\,\,\ep_{(01)(10)A}\ep_{(22)(21)A}=0\q\forall A\in H_3\\
  2.\q\q &\er^1=\left\{ \ep\in
\er(\es\cup\es^0)\,\,|\,\,\ep_{(01)(10)}\neq 0,\,\ep_{(10)(11)}\neq
0,\,\ep_{(01)(22)}\neq 0 \right\}\\
&\es^1:\,\,\ep_{(01)(10)A}\ep_{(10)(11)A}\ep_{(01)(22)A}=0\q\forall A\in H_3\\
  3.\q\q &\er^2=\left\{ \ep\in
\er(\es\cup\es^0\cup\es^1)\,\,|\,\,\ep_{(01)(10)}\neq 0,\,\ep_{(10)(11)}\neq 0 \right\}\\
&\es^2:\,\,\ep_{(01)(10)A}\ep_{(10)(11)A}=0\q\forall A\in H_3\\
  4.\q\q &\er^3=\left\{ \ep\in
\er(\es\cup\es^0\cup\es^1\cup\es^2)\,\,|\,\,\ep_{(01)(10)}\neq 0,\,\ep_{(02)(22)}\neq 0
\right\}\\ &\es^3:\,\,\ep_{(01)(10)A}\ep_{(02)(22)A}=0\q\forall A\in H_3\\
  5.\q\q &\er^4=\left\{ \ep\in
\er(\es\cup\es^0\cup\es^1\cup\es^2\cup\es^3)\,\,|\,\,\ep_{(01)(10)}\neq 0 \right\}\\
&\es^4:\,\,\ep_{(01)(10)A}=0\q\forall A\in H_3
\end{align*}

{\bf Step 1.}
In the rest of this subsection the parameters $a,b,c,\dots$ are
arbitrary complex numbers. Explicit solution under the assumption
$\ep_{(01)(10)}\neq 0$, $\ep_{(22)(21)}\neq 0$ can be written as
four parametric matrices. These matrices in $\er^0$ can be
equivalently replaced by renormalized matrices
$\er^0_{nor}=\{\ep^0_1,\ep^0_2,\ep^0_3,\ep^0_4\}$
where
\begin{alignat*}{2}
\ep^0_1 &=  \left( {\begin{smallmatrix}
0 & 0 & 1 & 0 & 0 & 1 & a & b \\
0 & 0 & 0 & 0 & 0 & 0 & c & d \\
1 & 0 & 0 & 0 & 0 & 0 & 0 & 0 \\
0 & 0 & 0 & 0 & 0 & 0 & 0 & 1 \\
0 & 0 & 0 & 0 & 0 & 0 & 0 & 0 \\
1 & 0 & 0 & 0 & 0 & 0 & 0 & 1 \\
a & c & 0 & 0 & 0 & 0 & 0 & 0 \\
b & d & 0 & 1 & 0 & 1 & 0 & 0
\end{smallmatrix}} \right), &\ \
\ep^0_2 &=  \left( {\begin{smallmatrix} 0 & 0 & 1 & d & ad & 1 & 1 & a \\ 0 & 0 & bad &
bad & bad & bd & b & ab \\ 1 & bad & 0 & 0 & dcba & c & cba & cba \\ d & bad & 0 & 0 &
dcba & d & c & 1 \\ ad & bad & dcba & dcba & & 0 & cba & ac \\ 1 & bd & c & d & 0 & 0 &
cb & 1 \\ 1 & b & cba & c & cba & cb & 0 & 0 \\ a & ab & cba & 1 & ac & 1 & 0 & 0
\end{smallmatrix}} \right),\\
\ep^0_3 &=  \left( {\begin{smallmatrix}
0 & 0 & 1 & 0 & 0 & 1 & 0 & 0 \\
0 & 0 & bd & 0 & 0 & ad & a & b \\
1 & bd & 0 & 0 & 0 & c & 0 & cb \\
0 & 0 & 0 & 0 & 0 & d & 0 & 1 \\
0 & 0 & 0 & 0 & 0 & 0 & 0 & 0 \\
1 & ad & c & d & 0 & 0 & ac & 1 \\
0 & a & 0 & 0 & 0 & ac & 0 & 0 \\
0 & b & cb & 1 & 0 & 1 & 0 & 0
\end{smallmatrix}} \right), &\ \
\ep^0_4 &=  \left( {\begin{smallmatrix} 0 & 0 & 1 & 0 & ad & 1 & 0 & a \\ 0 & 0 & bd & 0
& 0 & 0 & 0 & b\\ 1 & bd & 0 & 0 & 0 & c & 0 & cb \\ 0 & 0 & 0 & 0 & 0 & d & 0 & 1 \\ ad
& 0 & 0 & 0 & 0 & 0 & 0 & ac \\ 1 & 0 & c & d & 0 & 0 & 0 & 1 \\ 0 & 0 & 0 & 0 & 0 & 0 &
0 & 0 \\ a & b & cb & 1 & ac & 1 & 0 & 0
\end{smallmatrix}}
\right).
\end{alignat*}

{\bf Step 2.}
Note that the system of 48 equations $\es^0$ together with
$\ep_{(01)(10)}\neq 0$ enforces zeros on all positions from $\I_1^k$.
Moreover, the assumption $\ep_{(10)(11)}\neq 0$ and $\es^0$ enforces
further 4 zeros. Then the assumption $\ep_{(01)(10)}\neq
0$, $\ep_{(10)(11)}\neq 0$, $\ep_{(01)(22)}\neq 0$ gives us a single
solution:
$$
\er^1_{nor}=\{\ep^1\}, \q\q \text{where}\q
\ep^1 = \left( {\begin{smallmatrix} 0
& 0 & 1 & a & b & 1 & 0 & 0 \\ 0 & 0 & c & 0 & 0 & d & 0 & 0 \\ 1
& c & 0 & 0 & 1 & e & 0 & 0 \\ a & 0 & 0 & 0 & 0 & f & 0 & 0 \\ b
& 0 & 1 & 0 & 0 & 0 & 0 & 0 \\ 1 & d & e & f & 0 & 0 & 0 & 0 \\ 0
& 0 & 0 & 0 & 0 & 0 & 0 & 0 \\ 0 & 0 & 0 & 0 & 0 & 0 & 0 & 0
\end{smallmatrix}} \right)
$$

{\bf Step 3.}
Further solutions with assumption $\ep_{(01)(10)}\neq
0,\,\ep_{(10)(11)}\neq 0$, inequivalent to those in $\er^1$ and
$\er^0$, are listed below:
$$
\er^2_{nor}=\{\ep^2_1,\ep^2_2,\ep^2_3,\ep^2_4,\ep^2_5,\ep^2_6,
\ep^2_7,\ep^2_8\}
$$
\begin{alignat*}{3}
\ep^2_1 &=  \left( {\begin{smallmatrix}
0 & 0 & 1 & a & 0 & 0 & 0
& 0 \\ 0 & 0 & 0 & 0 & 0 & 0 & 0 & 0 \\ 1 & 0 & 0 & 0 & 1 & b & 0
& c \\ a & 0 & 0 & 0 & ac & d & e & ac \\ 0 & 0 & 1 & ac & 0
& 0 & 0 & 0 \\ 0 & 0 & b & d & 0 & 0 & 0 & 0 \\ 0 & 0 & 0 & e & 0
& 0 & 0 & 0 \\ 0 & 0 & c & ac & 0 & 0 & 0 & 0
\end{smallmatrix}} \right),&\
\ep^2_2 &=  \left( {\begin{smallmatrix}
 0 & 0 & 1 & a & 0 & 0 & 0 &
0 \\ 0 & 0 & 0 & 0 & 0 & 0 & 0 & 0 \\ 1 & 0 & 0 & 0 & 1 & b & 1 &
d \\ a & 0 & 0 & 0 & ad & 0 & c & ad \\ 0 & 0 & 1 & ad & 0 &
0 & 0 & 0 \\ 0 & 0 & b & 0 & 0 & 0 & 0 & 0 \\ 0 & 0 & 1 & c & 0 &
0 & 0 & 0 \\ 0 & 0 & d & ad & 0 & 0 & 0 & 0
\end{smallmatrix}} \right),&\
\ep^2_3 &=  \left( {\begin{smallmatrix}
0 & 0 & 1 & a & 0 & 0 & 0 &
0 \\ 0 & 0 & 1 & 0 & 0 & 0 & 0 & 0 \\ 1 & 1 & 0 & 0 & 1 & b & 0 &
c \\ a & 0 & 0 & 0 & ac & d & 0 & ac \\ 0 & 0 & 1 & ac & 0 &
0 & 0 & 0 \\ 0 & 0 & b & d & 0 & 0 & 0 & 0 \\ 0 & 0 & 0 & 0 & 0 &
0 & 0 & 0 \\ 0 & 0 & c & ac & 0 & 0 & 0 & 0
\end{smallmatrix}} \right),\\
\ep^2_4 &=  \left( {\begin{smallmatrix}
0 & 0 & 1 & a & 0 & 0 & 0 &
0 \\ 0 & 0 & 1 & 0 & 0 & 0 & 0 & 0 \\ 1 & 1 & 0 & 0 & 1 & b & 1 &
c \\ a & 0 & 0 & 0 & ac & 0 & 0 & ac \\ 0 & 0 & 1 & ac & 0 &
0 & 0 & 0 \\ 0 & 0 & b & 0 & 0 & 0 & 0 & 0 \\ 0 & 0 & 1 & 0 & 0 &
0 & 0 & 0 \\ 0 & 0 & c & ac & 0 & 0 & 0 & 0
\end{smallmatrix}} \right),&\ \
\ep^2_5 &=  \left( {\begin{smallmatrix} 0 & 0 & 1 & a & 0 & 0 & 0 &
0 \\ 0 & 0 & 1 & 1 & 0 & 0 & 0 & 0 \\ 1 & 1 & 0 & 0 & 1 & bd & b
& c \\ a & 1 & 0 & 0 & ac & d & bd & ac \\ 0 & 0 & 1 & ac
& 0 & 0 & 0 & 0 \\ 0 & 0 & bd & d & 0 & 0 & 0 & 0 \\ 0 & 0 & b &
bd & 0 & 0 & 0 & 0 \\ 0 & 0 & c & ac & 0 & 0 & 0 & 0
\end{smallmatrix}} \right),&\ \
\ep^2_6 &=  \left( {\begin{smallmatrix} 0 & 0 & 1 & a & 0 & 0 & 0 &
0 \\ 0 & 0 & 0 & 1 & 0 & 0 & 0 & 0 \\ 1 & 0 & 0 & 0 & 1 & 0 & 0 &
b \\ a & 1 & 0 & 0 & ab & c & d & ab \\ 0 & 0 & 1 & ab & 0 &
0 & 0 & 0 \\ 0 & 0 & 0 & c & 0 & 0 & 0 & 0 \\ 0 & 0 & 0 & d & 0 &
0 & 0 & 0 \\ 0 & 0 & b & ab & 0 & 0 & 0 & 0
\end{smallmatrix}} \right),\\
\ep^2_7 &=  \left( {\begin{smallmatrix} 0 & 0 & 1 & a & 0 & 0 & 0 &
0 \\ 0 & 0 & 0 & 1 & 0 & 0 & 0 & 0 \\ 1 & 0 & 0 & 0 & 1 & 0 & 1 &
b \\ a & 1 & 0 & 0 & ab & 0 & c & ab \\ 0 & 0 & 1 & ab & 0 &
0 & 0 & 0 \\ 0 & 0 & 0 & 0 & 0 & 0 & 0 & 0 \\ 0 & 0 & 1 & c & 0 &
0 & 0 & 0 \\ 0 & 0 & b & ab & 0 & 0 & 0 & 0
\end{smallmatrix}} \right),&\
\ep^2_8 &=  \left( {\begin{smallmatrix} 0 & 0 & 1 & a & b & 0 & 0 & 0 \\ 0 & 0 & c & d &
0 & e & 0 & 0 \\ 1 & c & 0 & 0 & 1 & 0 & 0 & 0 \\ a & d & 0 & 0 & 0 & f & 0 & 0 \\ b & 0
& 1 & 0 & 0 & 0 & 0 & 0 \\ 0 & e & 0 & f & 0 & 0 & 0 & 0 \\ 0 & 0 & 0 & 0 & 0 & 0 & 0 & 0
\\ 0 & 0 & 0 & 0 & 0 & 0 & 0 & 0
\end{smallmatrix}}
 \right).
\end{alignat*}

{\bf Step 4.} Now we can of course ignore the equations $\es^1$
because they are satisfied identically due to the system
$\es^2$. We list the next set
$$
\er^3_{nor}=\{\ep^3_1,\ep^3_2,\ep^3_3,\ep^3_4\}
$$
\begin{alignat*}{2}
\ep^3_1 &=\left( {\begin{smallmatrix}
0 & 0 & 1 & a & 0 & 0 & 0 & 0 \\ 0 & 0
& 0 & 0 & 0 & 1 & 1 & 0 \\ 1 & 0 & 0 & 0 & 0 & 0 & 0 & 0 \\ a & 0
& 0 & 0 & 0 & 0 & 0 & 0 \\ 0 & 0 & 0 & 0 & 0 & 0 & 0 & 0 \\ 0 & 1
& 0 & 0 & 0 & 0 & 0 & 0 \\ 0 & 1 & 0 & 0 & 0 & 0 & 0 & 0 \\ 0 & 0
& 0 & 0 & 0 & 0 & 0 & 0
\end{smallmatrix}} \right),&\q
\ep^3_2 &=  \left( {\begin{smallmatrix}
0 & 0 & 1 & a & 0 & 0 & 0 &
0 \\ 0 & 0 & b & 0 & 0 & 1 & 0 & 0 \\ 1 & b & 0 & 0 & 0 & 0 & 0 &
0 \\ a & 0 & 0 & 0 & 0 & c & 0 & 0 \\ 0 & 0 & 0 & 0 & 0 & 0 & 0 &
0 \\ 0 & 1 & 0 & c & 0 & 0 & 0 & 0 \\ 0 & 0 & 0 & 0 & 0 & 0 & 0 &
0 \\ 0 & 0 & 0 & 0 & 0 & 0 & 0 & 0
\end{smallmatrix}} \right),\\
\ep^3_3 &=  \left( {\begin{smallmatrix}
0 & 0 & 1 & 0 & 0 & 1 & 0 &
0 \\ 0 & 0 & 0 & 0 & 0 & 1 & 1 & 0 \\ 1 & 0 & 0 & 0 & 0 & 0 & 0 &
0 \\ 0 & 0 & 0 & 0 & 0 & 0 & 0 & 0 \\ 0 & 0 & 0 & 0 & 0 & 0 & 0 &
0 \\ 1 & 1 & 0 & 0 & 0 & 0 & 0 & 0 \\ 0 & 1 & 0 & 0 & 0 & 0 & 0 &
0 \\ 0 & 0 & 0 & 0 & 0 & 0 & 0 & 0
\end{smallmatrix}} \right),&\q
\ep^3_4 &=  \left( {\begin{smallmatrix}
0 & 0 & 1 & 0 & 0 & 1 & 0 &
0 \\ 0 & 0 & a & 0 & 0 & 1 & 0 & 0 \\ 1 & a & 0 & 0 & 0 & 0 & 0 &
0 \\ 0 & 0 & 0 & 0 & 0 & 0 & 0 & 0 \\ 0 & 0 & 0 & 0 & 0 & 0 & 0 &
0 \\ 1 & 1 & 0 & 0 & 0 & 0 & 0 & 0 \\ 0 & 0 & 0 & 0 & 0 & 0 & 0 &
0 \\ 0 & 0 & 0 & 0 & 0 & 0 & 0 & 0
\end{smallmatrix}} \right).
\end{alignat*}

{\bf Step 5.}
The systems $\es^0,\,\es^2,\,\es^3$ together with
$\ep_{(01)(10)}\neq 0$ give us 12 zeros and further 20 non-trivial
conditions. Adding two zeros following from $\es$ we obtain 3
solutions:
$$
\er^4_{nor}=\{\ep^4_1,\ep^4_2,\ep^4_3\}
$$
$$
\ep^4_1 =\left( {\begin{smallmatrix}
0 & 0 & 1 & 0 & 0 & 0 & 0 & a
\\ 0 & 0 & 0 & b & 0 & 0 & c & 0 \\ 1 & 0 & 0 & 0 & 0 & 0 & d & 0 \\
0 & b & 0 & 0 & 0 & 0 & 0 & e \\ 0 & 0 & 0 & 0 & 0 & 0 & 0 & 0 \\ 0
& 0 & 0 & 0 & 0 & 0 & 0 & 0 \\ 0 & c & d & 0 & 0 & 0 & 0 & 0 \\ a & 0
& 0 & e & 0 & 0 & 0 & 0
\end{smallmatrix}} \right),\q
\ep^4_2 =  \left( {\begin{smallmatrix}
0 & 0 & 1 & a & 0 & 0 & 0 &
0 \\ 0 & 0 & b & c & 0 & 0 & 0 & 0 \\ 1 & b & 0 & 0 & 0 & 0 & 0 &
0 \\ a & c & 0 & 0 & 0 & 0 & 0 & 0 \\ 0 & 0 & 0 & 0 & 0 & 0 & 0 &
0 \\ 0 & 0 & 0 & 0 & 0 & 0 & 0 & 0 \\ 0 & 0 & 0 & 0 & 0 & 0 & 0 &
0 \\ 0 & 0 & 0 & 0 & 0 & 0 & 0 & 0
\end{smallmatrix}} \right),\q
\ep^4_3 =  \left( {\begin{smallmatrix}
0 & 0 & 1 & 0 & 0 & a & 0 &
0 \\ 0 & 0 & 0 & 0 & 0 & 0 & 0 & 0 \\ 1 & 0 & 0 & 0 & 0 & b & 0 &
0 \\ 0 & 0 & 0 & 0 & 0 & 0 & 0 & 0 \\ 0 & 0 & 0 & 0 & 0 & 0 & 0 &
0 \\ a & 0 & b & 0 & 0 & 0 & 0 & 0 \\ 0 & 0 & 0 & 0 & 0 & 0 & 0 &
0 \\ 0 & 0 & 0 & 0 & 0 & 0 & 0 & 0
\end{smallmatrix}} \right).
$$

Since all pairs of relevant indices lie in {\it one} orbit (the
whole set $\I$) the system $\es^4:\ep_k=0,\,\forall k\in \I$
enforces zeros on all 24 positions. This precisely means that now
{\it only the trivial zero solution is inequivalent to solutions
in $\er^0,\,\er^1,\,\er^2,\,\er^3,\,\er^4$, i.e. we have evaluated
the whole $\er(\es)$ up to equivalence.}

Our final goal in this part. is to compute the complete set of
inequivalent {\it normalized} solutions. Thus in the final stage
we took each solution matrix and discussed all possible
combinations of zero or non-zero parameters like in the following
example.

\begin{exmp}\label{nonzer}
For instance take the matrix $\ep^0_2$ and let all its parameters
be non-vanishing. Our question is whether or not it is possible to
normalize it to the trivial contraction matrix $\ep_0$ which has
all relevant epsilons equal to unity. Then the resulting graded
contractions would be isomorphic to the algebra $sl(3,\C)$ for
arbitrary non-zero values of parameters in $\ep^0_2$. We have
verified that the system of 24 equations corresponding to the
matrix equality $\ep^0_2\bullet\ \alpha=\ep_0$ has a general
solution in $\C\backslash\{0\}$. The matrix $\ep^0_2$ with
non--zero parameters is then equivalent to the trivial solution
$\ep_0$ and the corresponding graded contraction is isomorphic to
$sl(3,\C)$.
\end{exmp}

Similar calculations had to be done for all matrices
$\ep^0_1,\dots,\ep^4_3$. Results are given in the Appendix.


\section{Algorithm of identification}
The goal of our work is to determine the structure of Lie algebras
from the structure constants. Each of the 188 contraction matrices
found there uniquely determines a set of structure constants of an
8-dimensional Lie algebra. Unfortunately existing methods
(\cite{RWZ,BKS} and references cited therein) do not allow to
recognize all 88 solvable indecomposable algebras found here among
the 188 cases. Therefore practical necessity of this paper was to
develop further the identification algorithm.

In this section, after recalling pertinent properties of Lie
algebras, we describe seven steps of our algorithm.

  Let
$\mL$ be a complex Lie algebra of dimension $n$. We denote the
{\bf derived algebra} of $\mL$ by $$ D(\mL)=[\mL,\mL], $$ and the
{\bf center} of $\mL$ by $$ C(\mL)=\{x\in \mL \mid\forall y \in
\mL,\ [x,y]=0\}. $$
  If $\mL_1, \mL_2\subset \mL$ are subalgebras of $\mL$, such that
$\mL = \mL_1 +\mL_2 $ (as vector spaces), then the Lie algebra
$\mL$ is:
\begin{enumerate}
\item
the direct sum of its ideals $\mL_1, \mL_2$, denoted $\mL = \mL_1
\oplus \mL_2$, if $$
 [\mL_1,\mL_2]=0,\qquad [\mL_i,\mL_i]\subseteq\mL_i,\qquad    i=1,2,
$$
\item
the semidirect sum of its ideal $\mL_1$ and subalgebra
    $\mL_2$, denoted $\mL = \mL_1  \in\!\!\!\!\!\! + \ \mL_2$, if
$$
 [\mL_1,\mL_2]\subseteq \mL_1,\qquad [\mL_i,\mL_i]\subseteq\mL_i,\qquad
    i=1,2.
$$
\end{enumerate}

We now describe the method of identification in particular steps.

\noindent {\bf Step 1. Splitting of the maximal central
component.} \

If the complement $X = C(\mL)\backslash D(\mL)$ of the derived
algebra to the center is non-empty, then the central decomposition
of the Lie algebra $\mL$ can be obtained from the decomposition of
the quotient algebra
\begin{equation}
\mL/D(\mL) = X/D(\mL)\oplus \widetilde{\mL}/D(\mL),
\end{equation}
where $X/D(\mL)$ denotes the set of all cosets modulo $D(\mL)$
that are represented by the elements of $X$ and
$\widetilde{\mL}/D(\mL)$ is the complement of $X/D(\mL)$ in
$\mL/D(\mL)$, where $\widetilde{\mL}$ contains $D(\mL)$. From now
on we consider Lie algebras without a separable central component.
For those algebras $C(\mL)\subseteq D(\mL)$ holds.

\noindent {\bf Step 2. Decomposition into a direct sum of
indecomposable ideals.}

 Let
\begin{equation}
C_R(\ad(\mL)) = \{x\in R \mid \forall y \in \ad(\mL),\ [x,y]=0 \}
\end{equation}
denote the centralizer of the adjoint representation of $\mL$ in
the ring $R=\bC^{n,n}$. An idempotent $E\in R$ in the ring $R$ is
a nonzero matrix satisfying $E^2=E$. Lie algebra $\mL$ is
decomposable into a direct sum of its ideals if and only if there
exists a non--trivial idempotent $1 \neq E \in C_R(\ad(\mL))$. In
such a case the decomposition has the form $$
 \mL=\mL_0\oplus \mL_1,\qquad [\mL_0,\mL_1]=0,\qquad
[\mL_i,\mL_i]\subseteq \mL_i,\qquad i=0,1, $$ where $\mL_0,\mL_1$
are eigen-subspaces of the idempotent $E$ corresponding to the
eigenvalues 0,1. In this step, all decomposable algebras are
decomposed and from now on we consider indecomposable algebras
only. For an explicit algorithm of finding a nontrivial idempotent
see \cite{RWZ}.

\noindent {\bf Step 3. Determination of the radical, the Levi
decomposition and the nilradical.}

First we find the radicals for all algebras. The {\bf radical}
$R(\mL)$ of the Lie algebra $\mL$ is the maximal solvable ideal in
$\mL$ and in our case
\begin{equation}
R(\mL) = \{x\in \mL \mid \forall y \in D(\mL), \
\operatorname{Tr}(\ad(x)\ad(y))=0 \}.
\end{equation}
Then we use Levi's theorem which states that for an arbitrary
finite dimensional Lie algebra $\mL$ over a field of
characteristic zero there exists a semisimple subalgebra
$\mathcal{S}$ such that
\begin{equation}
\mL = R(\mL) \in\!\!\!\!\!\! + \  \mathcal{S} ,
\end{equation}
and we make the semidirect decomposition of all Lie algebras. If
$R(\mL)=0$, the Lie algebra $\mL$ is semisimple and if
$\mathcal{S}=0$, it is solvable. At last we find the maximal
nilpotent ideals ({\bf nilradicals}) from radicals of all algebras
and determine which algebras are nilpotent. The algorithm for
providing Levi's decompositions and algorithm of computing the
nilradical are completely described in \cite{RWZ}.

\noindent {\bf Step 4. Computation of the derived series, the
lower central series and the upper central series.}

The {\bf derived series} of Lie algebra $\mL$ is a series of
ideals $D^0(\mL) \supseteq D^1(\mL) \supseteq D^2(\mL) \supseteq
\ldots \supseteq D^k(\mL) \supseteq \ldots$ defined by: $$
 D^0(\mL) = \mL, \qquad D^{k+1}(\mL) = [D^k(\mL),D^k(\mL)],
\qquad k=0,1,2,\ldots. $$ The {\bf lower central series} of $\mL$
is a series of ideals $\mathcal{L}^0 \supseteq \mathcal{L}^1
\supseteq \mathcal{L}^2 \supseteq \ldots \supseteq \mathcal{L}^k
\supseteq \ldots$ defined by: $$ \mL^0 = \mL, \qquad \mL^{k+1}
=[\mL^k,\mL], \qquad k=0,1,2,\ldots. $$ The {\bf upper central
series} of $\mL$ is a series of ideals $C^0(\mL) \subseteq
C^1(\mL) \subseteq C^2(\mL) \subseteq \ldots \subseteq C^k(\mL)
\subseteq \ldots $ defined by: $$ C^0(\mL) = 0, \qquad
C^{k+1}(\mL)/C^k(\mL)=C(\mL/C^k(\mL)), \qquad k=0,1,2,\ldots. $$
The dimensions of ideals in the above series are invariants of
$\mL$. In Step 4 we divide all Lie algebras into classes according
to these invariants.

\noindent {\bf Step 5. Determination of the algebra of
derivations.}

A linear mapping $d:\mL\rightarrow\mL$ is called a {\bf
derivation} of the Lie algebra $\mL$ if
\begin{equation}
d[x,y] = [d x,y] + [x,d y], \qquad \forall x,y\in\mL.
\end{equation}
The set of all derivations of $\mL$ forms a Lie algebra
$\Der(\mL)\subseteq gl(\mL)$ called {\bf Lie algebra of
derivations} of $\mL$. If $c_{i,j}^k$ are the structure constants
of $\mL$ in the basis $\{e_i\}_{i=1}^n$ and let $d e_i
=\sum_{j=1}^n d_{ji}e_j,\ \forall i\in\hat{n}= \{1,2,\ldots,n\}$,
then $d\in\Der(\mL)$, if
\begin{equation}
\sum_{m=1}^n(c_{ij}^m d_{km} - c_{mj}^k d_{mi} - c_{im}^k d_{mj})
=0, \qquad \forall i,j,k\in\hat{n}.
\end{equation}
The dimension of the algebra of derivations is an invariant for
the Lie algebra $\mL$, and we divide each class from the previous
step into new classes according to $\dim(\Der(\mL))$.

\noindent {\bf Step 6. Casimir operators.}

An element $F$ of the universal enveloping algebra of $\mL$ which
satisfies
\begin{equation}\label{deca}
[x,F] = 0, \qquad \forall x\in \mL,
\end{equation}
is called Casimir operator \cite{AA,PSWZ}. These operators can be
calculated as follows. Represent the elements of basis
$\{e_i\}_{i=1}^n$ of $\mL$ by the vector fields
\begin{equation}
e_i \rightarrow\hat{x}_i =
\sum_{j,k=1}^nc_{ij}^kx_k\frac{\partial}{\partial x_j},
\end{equation}
which have the same commutation rules and act on the space of
continuously differentiable functions $F(x_1,\ldots,x_n)$. A
function $F$ is called {\bf formal invariant} of $\mL$ if it is a
solution of
\begin{equation}
\hat{x}_iF=0,\qquad i\in\hat{n}.
\end{equation}
The number of algebraically independent formal invariants is
\begin{equation}
\tau(\mL) = \dim(\mL)-r(\mL),
\end{equation}
where $r(\mL)$ is the rank of the antisymmetric matrix $M_{\mL}$
with entries $(M_{\mL})_{ij} = \sum_{k}c_{ij}^ke_k$:
\begin{equation}
r(\mL) =
\begin{array}[t]{c}
  \sup \\
  ^{(e_1,\ldots,e_n)} \\
\end{array} \operatorname{rank}(M_{\mL}).
\end{equation}
From a polynomial formal invariant $F(x_1,\ldots,x_n)$ the Casimir
invariant is obtained according to the recipe: replace the
variables $x_i$ by noncommuting elements $e_i$ and symmetrize
$F(e_1,\ldots,e_n)$ \cite{AA}. In the case of nilpotent algebras
all solutions $F(x_1,\ldots,x_n)$ can be written as polynomial
functions yielding Casimir operators. We can decide whether two
algebras are non-isomorphic by comparing the numbers of their
formal invariants or, in the case of polynomial invariants,
according to their order.

\noindent {\bf Step 7. Seeking isomorphisms in the classes of
algebras.}

Complex Lie algebras $\mL$ and $\mL'$  of the same dimension $n$
determined by structure constants $x_{ij}^k$ and $y_{ij}^k$ are
isomorphic if there exists a regular matrix $A\in\bC^{n,n}$, whose
elements satisfy the system of $n^2(n-1)/2$ quadratic equations
\begin{equation}\label{ire}
\sum_{r=1}^{n} x_{ij}^{r}A_{kr} = \sum_{\mu, \nu =1}^{n} A_{\mu
i}A_{\nu j}y_{\mu\nu}^k, \qquad i=1,\ldots,n-1,\quad j=
i,\ldots,n,\quad k\in\hat{n}.
\end{equation}
We can either solve this system on computer and test solutions on
regularity or try to prove that this system has no regular
solution.


\section{Results of the identification}
The set of  188 inequivalent solutions of the system of
contraction equations for the Pauli graded Lie algebra $sl(3,\bC)$
was divided into 13 groups according to the numbers $\nu$ of zeros
among 24 relevant entries in the contraction matrices
$\varepsilon$. These solutions are denoted $\varepsilon^{\nu,i}$,
where the second index $i$ is numbering solutions with the same
$\nu$. Correspondingly, the contracted Lie algebra corresponding
to solution $\varepsilon^{\nu,i}$ is denoted $\mL_{\nu,i}$. The
following table gives the numbers of solutions $\varepsilon$
corresponding to each $\nu$: $$
\small
\begin{tabular}{l||l|l|l|l|l|l|l|l|l|l|l|l|l}
$\nu$ & 0 & 9 & 12 & 15 & 16 & 17 & 18 & 19 & 20 & 21 & 22 & 23 &
24\\ \hline $1\leq i\leq $  & 1 & 1 & 2  & 7  & 7  & 17 & 36 & 45
& 42 & 21 & 7 &  1 & 1\\
\end{tabular}
$$

Among the 188 solutions there are two trivial solutions. One
trivial solution $\varepsilon^{24,1}$, with 24 zeros, corresponds
to the 8--dimensional Abelian Lie algebra while the other trivial
solution $\varepsilon^{0,1}$, without zeros, corresponds to the
initial Lie algebra $sl(3,\bC)$. From the remaining 186 nontrivial
solutions, 11 solutions depend on one non-zero complex parameter
$a$ and two depend on two non-zero complex parameters $a$, $b$.
The corresponding parametric algebras are denoted by
$\mL_{\nu,i}(a)$, $\mL_{\nu,i}(a,b)$.

In Step 1, 71 algebras allowed the separation of a central
component. For these algebras only non-Abelian parts with
dimension lower than 8 were further investigated. We will denote
the non-Abelian part of the Lie algebra $\mL$ by $\mL'$.

In Step 2 only 12 algebras were decomposable. Decomposition was in
all cases the direct sum of two indecomposable ideals. At this
stage we are left with 198 indecomposable algebras for the next
steps leading to their identification.

The Levi decomposition in Step 3 was trivial for all algebras in
the sense that there is no semisimple part. Of the resulting Lie
algebras 24 were solvable (non-nilpotent) and 174 nilpotent.

In Step 4, computation of dimensions of the derived series, the
lower central series and the upper central series improved the
accuracy of determination of the number of Lie algebras: solvable
algebras have split into 18 classes and nilpotent algebras into 52
classes, to be still refined in the following steps.

Dimensions of algebras of derivations in Step 5 divided nilpotent
algebras into 100 classes and retained 18 classes of solvable
algebras.

In Step 6 the numbers of formal invariants of Lie algebras and of
Casimir invariants for the nilpotent Lie algebras were computed.
It lead to 115 classes of nilpotent algebras. The Casimir
invariants were computed straightforwardly from the definition
\eqref{deca}.

In Step 7, computer computation discovered 4 isomorphic pairs
among the solvable algebras and 52 isomorphic pairs among the
nilpotent algebras. Furthermore, among 11 parametric algebras, two
solvable algebras and two nilpotent ones were extended to zero
values of parameters:
\begin{equation} \label{ext}
\begin{array}{ll}
\mL_{15,6}(0)\cong\mL_{16,6},
&\mL_{15,5}(a,0)\cong\mL_{16,2}(\frac{1}{a}), \\
\mL_{12,2}(0,b)\cong\mL_{15,7}(\frac{1}{b^2}), &
\mL_{18,25}(0)\cong\mL_{19,26}, \\
 \end{array}
\end{equation}
So we obtained 18 non-isomorphic (non-nilpotent) solvable
algebras. Moreover the structure of all decomposable algebras was
completed. We give the structure of mutually non-isomorphic
decomposable algebras:
\begin{equation} \label{decom}
\begin{array}{lll}
\mL_{18,32}=\mL'_{21,9}\oplus\mL'_{21,9},
  &\mL_{19,36}=\mL'_{21,9}\oplus\mL'_{22,1},
&\mL_{20,10}=\mL'_{23,1}\oplus\mL'_{21,2},\\
\mL'_{20,20}=\mL'_{23,1}\oplus\mL'_{21,9}, &
\mL_{20,21}=\mL'_{22,1}\oplus\mL'_{22,1},
  &\mL'_{21,4}=\mL'_{23,1}\oplus\mL'_{22,1},\\
\mL_{21,15}=\mL'_{23,1}\oplus\mL'_{22,3},
  &\mL'_{22,2}=\mL'_{23,1}\oplus\mL'_{23,1},\\
\end{array}
\end{equation}
as well as the list of all isomorphisms among non-decomposed
algebras: $$
\begin{array}{lll}
\mL'_{22,2}\cong\mL'_{22,6}, & \mL_{20,21}\cong\mL_{20,22}, &
\mL'_{21,4}\cong\mL'_{21,6}\cong\mL'_{21,10},\\
\mL'_{22,3}\cong\mL'_{22,4}\cong\mL'_{22,5}, &
\mL'_{21,3}\cong\mL'_{21,5}\cong\mL'_{21,8}, &
\mL'_{21,11}\cong\mL'_{21,13}\cong\mL'_{21,18},\\
\mL'_{21,12}\cong\mL'_{21,14}\cong\mL'_{21,17}, &
\mL'_{20,14}\cong\mL'_{20,23}\cong\mL'_{20,28}, &
\mL'_{20,19}\cong\mL'_{20,26}\cong\mL'_{20,29},\\
\mL'_{20,13}\cong\mL'_{20,15}\cong\mL'_{20,24}, &
\mL_{20,32}\cong\mL_{20,37}, & \mL_{20,31}\cong\mL_{20,34}, \\
\mL_{19,31}\cong\mL_{19,33}\cong\mL_{19,39}, &
\mL_{20,17}\cong\mL_{20,18}\cong\mL_{20,27}, &
\mL_{20,33}\cong\mL_{20,36}, \\
\mL_{19,25}\cong\mL_{19,38}\cong\mL_{19,40}, &
\mL_{20,12}\cong\mL_{20,25}\cong\mL_{20,30}, &
\mL_{18,26}\cong\mL_{18,31}, \\
\mL_{19,28}\cong\mL_{19,34}\cong\mL_{19,35}, &
\mL_{18,25}(a)\cong\mL_{18,30}(a)\cong\mL_{18,33}(a), &
\mL_{19,26}\cong\mL_{19,30}\cong\mL_{19,37}. \\
\end{array}
$$

\section{Algebra of derivations}
Computation of the set of invariants according to \cite{RWZ}
turned out to be insufficient for our purpose: in many cases there
were several Lie algebras with the same characteristics proposed
in \cite{RWZ}. Even if Casimir operators were used, we did not
attain unique characterization. Unexpectedly, additional
computation of the dimensions of algebras of derivations finally
almost solved the problem.

There remained still the following undetermined cases among the
nilpotent algebras:
\begin{equation}\label{uc}
\begin{array}{c}
  \mL_{17,2}\ $and$\ \mL_{19,22}, \qquad \mL_{17,13}(a) \ $and$\
\mL_{18,11}, \qquad \mL_{18,16} \ $and$\  \mL_{19,20}, \\
\\
  \mL_{15,5}(a,b) \ $and$\  \mL_{17,10}, \qquad \mL_{16,3}(a) \
$and$\ \mL_{17,9} \ $and$\  \mL_{17,12}. \\
\end{array}
\end{equation}

Determination of the algebra of derivations is a linear problem.
If two Lie algebras are isomorphic, then their algebras of
derivations must be isomorphic too. So we can apply our algorithm
of identification to algebras of derivations of Lie algebras
\eqref{uc}. Obtained characterizations of algebras of derivations
are given in Table \ref{tab1}.  All investigated algebras of
derivations were indecomposable and without separable central
component. The dimensions of their derived series, lower central
series and upper central series are given in the 2nd, 3rd and 4th
column. The numbers $\tau$ of formal invariants of algebras of
derivations are given in the 5th column. We constructed also the
{\bf sequences of algebras of derivations}
 \begin{equation}
 \Der^k(\mL) =
 \underbrace{\Der(\ldots(\Der(\mL)))}_{k-times},\qquad
 k\in \mathbb{N}.
 \end{equation}
Dimensions of members of these sequences are in the 6th column of
the Table \ref{tab1}. In the same way like for the series in the
2nd, 3rd and 4th column, we did not write out the repeating
numbers at the end of sequence.

\begin{table}[h!]
\small
\begin{tabular}{|l||l|l|l|l|l|}
\hline $\mL$ & $D^k(\Der(\mL))$ & $(\Der(\mL))^k$ &
$C^k(\Der(\mL))$ & $\tau$ & $\Der^k(\mL)$ \\ \hline \hline
$\mL_{17,2}$ & (17,15) & (17,15) & (0) & 3 & 17,19\\ \hline
$\mL_{19,22}$ & (17,14,8,0) & (17,14) & (0) & 3 & 17,19\\ \hline
$\mL_{15,5}(a,b),\  ^{a\neq 0}_{(-1,0)\neq (a,b) \neq (1,1)}$ &
(16,15,6,0) & (16,15) & (0) & 6 & 16\\ \hline $\mL_{17,10}$ &
(16,15,6,0) & (16,15) & (0) & 6 & 16,19\\ \hline $\mL_{17,13}(a),\
0\neq a\neq -1$ & (17,15,6,0) & (17,15) & (0) & 5 & 17\\ \hline
$\mL_{18,11}$ & (17,15,6,0) & (17,15) & (0) & 5 & 17,18\\ \hline
$\mL_{18,16}$ & (19,17,12,3,0) & (19,17) & (0) & 7 & 19\\ \hline
$\mL_{19,20}$ & (19,16,9,0) & (19,16) & (0) & 7 & 19,25\\ \hline
$\mL_{16,3}(a),\ a\neq 0$ & (16,15,6,0) & (16,15) & (0) & 6 & 16\\
\hline $\mL_{17,9}$ & (16,15,6,0) & (16,15) & (0) & 6 & 16\\
\hline $\mL_{17,12}$ & (16,15,6,0) & (16,15) & (0) & 6 &  16\\
\hline
\end{tabular}

\medskip

\centering \caption[l]{{\it Characterizations of algebras of
derivations: derived series, lower central series, upper central
series, numbers of formal invariants and sequences of algebras of
derivations.}} \vspace{-8pt}\label{tab1}
\end{table}

Table \ref{tab1} shows that there remains the unresolved case of
three nilpotent Lie algebras $\mL_{16,3}(a)$, $\mL_{17,9}$ and
$\mL_{17,12} $; other algebras are non-isomorphic. The nonzero
commutation relations of these Lie algebras are:
\begin{equation}
\begin{array}{ll}
\mL_{16,3}(a)\quad & [e_1,e_3]=-ae_5,\ [e_1,e_4]=e_8,\
[e_1,e_5]=e_7,\ [e_1,e_6]=e_4,\\
              &  [e_2,e_3]=e_7,\ [e_2,e_6]=e_8,\ [e_3,e_5]=e_8,\
[e_4,e_6]=e_7\\
\\
\mL_{17,9}\quad & [e_1,e_3]=e_5,\ [e_1,e_4]=e_8,\ [e_1,e_5]=e_7,\
[e_1,e_6]=e_4,\\
            & [e_2,e_3]=e_7,\ [e_3,e_5]=e_8,\ [e_4,e_6]=e_7 \\
\\
\mL_{17,12}\quad & [e_1,e_3]=e_5,\ [e_1,e_4]=e_8,\ [e_1,e_6]=e_4,\
[e_2,e_3]=e_7,\\
            & [e_2,e_6]=e_8,\ [e_3,e_5]=e_8,\ [e_4,e_6]=e_7 \\
\end{array}
\end{equation}

Let us first prove that Lie algebras $\mL_{17,9}$ and $
\mL_{17,12}$ are not isomorphic. The system \eqref{ire} has in
this case 140 equations. Some equations are of the form
$A_{i,j}=0$. After multiple use of these equations in the system
there remain only 40 equations and the matrix of isomorphism
$A\in\bC^{8,8}$ has the form
\begin{equation}
\left(%
\begin{smallmatrix}\label{mat}
A_{1,1} & A_{1,2} & A_{1,3} & 0 & 0 & A_{1,6} & 0 & 0 \\ A_{2,1} &
A_{2,2} & A_{2,3} & 0 & 0 & A_{2,6} & 0 & 0 \\ A_{3,1} & A_{3,2} &
A_{3,3} & 0 & 0 & A_{3,6} & 0 & 0 \\ A_{4,1} & A_{4,2} & A_{4,3} &
A_{4,4} & A_{4,5} & A_{4,6} & 0 & 0\\ A_{5,1} & A_{5,2} & A_{5,3}
& A_{5,4} & A_{5,5} & A_{5,6} & 0 & 0\\ A_{6,1} & A_{6,2} &
A_{6,3} & 0 & 0 & A_{6,6} & 0 & 0 \\ A_{7,1} & A_{7,2} & A_{7,3} &
A_{7,4} & A_{7,5} & A_{7,6} & A_{7,7} & A_{7,8}\\ A_{8,1} &
A_{8,2} & A_{8,3} & A_{8,4} & A_{8,5} & A_{8,6} & A_{8,7} &
A_{8,8}\\
\end{smallmatrix}%
\right).
\end{equation}
The simplest equations of the system are
\begin{equation}
\begin{array}{c}\label{rce1}
A_{6,2}A_{4,4}=0, \quad A_{6,2}A_{4,5}=0, \quad A_{6,3}A_{4,4}=0,
\quad A_{6,6}A_{4,5}=0, \\
\\
A_{6,1}A_{4,4}=A_{7,8}, \quad A_{6,1}A_{4,5}=A_{7,7}, \quad
A_{6,3}A_{4,5}=A_{7,8},\quad A_{6,6}A_{4,4}=A_{7,7}.\\
\end{array}
\end{equation}
If we put $A_{4,4}=0$ then $A_{7,7}=0,\ A_{7,8}=0$ and the matrix
$A$ is singular. On the other hand  $A_{4,4}\neq 0$ implies
$A_{6,3}=0$ and consequently $A_{7,8}=0$. From $A_{4,4}\neq 0$ and
$A_{7,8}=0$ now follows $A_{6,1}=0$ and consequently $A_{7,7}=0$.
Thus the matrix $A$ is again singular and the proof is completed.

In the remaining cases $\mL_{16,3}(a)$ {\it vs.} $\mL_{17,9}$ and
$\mL_{16,3}(a)$ {\it vs.} $\mL_{17,12}$ the matrices of
isomorphism have the same form as \eqref{mat}. Moreover in the
case $\mL_{16,3}(a)$ {\it vs.} $\mL_{17,12}$ there are the
equations \eqref{rce1} in system \eqref{ire} and therefore the Lie
algebras $\mL_{16,3}(a)$ and $\mL_{17,12}$ are non-isomorphic. The
proof that the Lie algebras $\mL_{16,3}(a)$ and $\mL_{17,9}$ are
non-isomorphic is similar to the first one, but more complicated
and we shall not describe it there. This finished the
identification.

Table \ref{tab2} summarizes our results. For each dimension of
non-Abelian part of Lie algebra in the first column, the numbers
of obtained Lie algebras are given in the other columns according
to their types.
\begin{table}[h!]
\begin{tabular}{|c||c|c|c|c|c|}
\hline Dimension of& \multicolumn{2}{c|}{Solvable} &
\multicolumn{2}{c|}{Nilpotent} & Total \\ \cline{2-5} non--Abelian
part & Indecomp. & Decomp. & Indecomp. & Decomp. &
\\
\hline \hline  3 &  &  & 1 &  & 1\\ \hline  4 & 1 &  & 1 &  & 2\\
\hline  5 & 1 &  & 4 &  & 5\\ \hline  6 & 1 &  & 9 & 1 & 11\\
\hline  7 & 4 & 1 & 28 & 1 & 34\\ \hline  8 & 11 & 2 & 77 & 3 &
93\\ \hline
\end{tabular}

\medskip \caption{\emph{The numbers of contracted Lie
algebras from Pauli graded $sl(3,\bC)$.}}\label{tab2}
\end{table}
Together with the 8-dimensional Abelian Lie algebra and the Lie
algebra $sl(3,\bC)$ the final number of all contracted Lie
algebras is 148.


\section{Concluding remarks}
\begin{itemize}

\item  Using the symmetry group of the Pauli grading of $sl(3,\C)$,
we have evaluated the set of all 188 solutions of the
corresponding contraction system up to equivalence. For the
solution of the normalization equations and for the explicit
evaluation of orbits of solutions we used the computer program
MAPLE 8. A new method of solving is based on Theorem \ref{main}.
It enabled us to check all solutions in the sets $\er^0$, $\er^1$,
$\er^2$, $\er^3$ and $\er^4$ also by hands. It is interesting to
note that $$ \mbox{min} \set{\nu(\ep)\in\{1,2,3,\dots\}~\big|~\ep
\in \er(\es)}=9 $$ i.e. there are no solutions with less than 9
zeros (besides the trivial solution $\ep_0$). Moreover, there are
no solutions with 10, 11, 13 or 14 zeros. The complete list of
solutions is put in the Appendix. It serves as an input for
further analysis --- the identification of resulting Lie algebras.

\item Method used for identification of resulting Lie algebras
was described in Section 7 and 9. For all steps of this method a
program in MAPLE~8 was written and applied, in the given order, to
each contraction matrix. The number of all non-isomorphic
contracted algebras is 148.

\item In physical applications, only contractions of the continuous
kind are often considered. From the point of view of our method it
is only a part of the solution of the contraction system. We have
studied this question and the continuous or discrete type is
distinguished by the corresponding subscript of each contraction
matrix given in the Appendix.

\item Our experience shows that a simple minded application of even
powerful symbolic languages does not provide all solutions. The
second crucial point is the introduction of equivalence classes of
solutions which are difficult to enforce in the symbolic
calculation.

\item Our results can be compared with graded contractions for the
root decomposition \cite{ALPW}. Their investigations resulted in
32 Lie algebras, of which 9 are 8-dimensional non-decomposable.

\item Apparently the possibility of three-term contraction
equations has not been noticed in the literature before. They may
appear for finest gradings, i.e. when the grading subspaces are
one-dimensional. Our case of $sl(3,\C)$ is too low and all
contraction equations reduce to two-term ones. Following example
shows that three--term contraction equations may arise for the
Pauli grading of $sl(5,\C)$. Let us write the contraction equation
$e((01)(10)(31))$. Since the subspaces are one-dimensional, we
have $$
 [X_{01},[X_{10},X_{31}]_\ep]_\ep
                      +[X_{31},[X_{01},X_{10}]_\ep]_\ep
                      +[X_{10},[X_{31},X_{01}]_\ep]_\ep =0.
$$ Using (\ref{komutator}), we obtain a three-term equation
\begin{align*}
 &\ep_{(01)(10)}\ep_{(11)(31)}(\omega_5-1)(\omega_5^3-\omega_5)
+\ep_{(10)(31)}\ep_{(41)(01)}(1-\omega_5)(1-\omega_5^4)\\
+&\ep_{(31)(01)}\ep_{(32)(10)}(1-\omega_5^3)(\omega_5^2-1)=0,
\end{align*}
where $\omega_5$ is the fifth root of unity.

\item The ranges of continuous parameters in the 13 contraction
matrices where they appear, still need to be restricted by
 the requirement that for every two values of a parameter
 one has non-isomorphic Lie algebras.

\item
The resulting number of non-isomorphic parametric Lie algebras
given by one- or two-parameter solutions is 11. They would deserve
further study in order to determine the ranges of the parameters.
Allowing for zero values of complex parameters $a$, $b$ we
obtained the following isomorphisms: $$
\begin{array}{lll}
\mL_{12,2}(0,0) = \mL_{18,28}, & \mL_{12,2}(a,0) \cong
\mL_{15,6}(a^2), & \mL_{12,2}(0,b) \cong
\mL_{15,7}(\frac{1}{b^2}), \\ \mL_{15,5}(0,0) \cong \mL_{17,9}, &
\mL_{15,5}(a,0) \cong \mL_{16,2}(\frac{1}{a}), & \mL_{15,5}(0,b)
\cong \mL_{16,3}(b), \\ \mL_{15,6}(0) = \mL_{16,6}, &
\mL_{15,7}(0) \cong \mL_{16,6}, & \mL_{16,1}(0) = \mL_{17,3}, \\
\mL_{16,2}(0) \cong \mL_{17,12}, & \mL_{16,3}(0) \cong
\mL_{17,14}, & \mL_{17,7}(0) \cong \mL_{18,2},\\ \mL_{17,13}(0)
\cong \mL_{18,12}, & \mL_{18,29}(0) \cong \mL_{19,29}, &
\mL_{18,25}(0) = \mL_{19,26}.\\
\end{array}
$$ Here, as elsewhere, $\cong$ denotes isomorphism and $=$
identity
 of Lie algebras, i.e. the commutation relations
are the
 same. We made the extensions to zero values of
parameters, whenever the derived series, the lower central series
and the
 upper central series were the same for nonzero and zero values of
 parameters \eqref{ext}.
However, we were not able to determine exact ranges

of parameters except for the cases of

$\mL_{16,3}(a), \mL_{17,13}(a), \mL_{18,25}(a), \mL_{18,29}(a)$,
where the range of complex parameter $a\neq 0$ could be restricted
to $0 < |a| \leq 1$.

\item Let us compare our results with results of \cite{ALPW},
where the contractions of $sl(3,\bC)$ graded by the maximal torus
were studied. The toroidal and Pauli gradings of $sl(3,\bC)$ share
one common coarser $\mathbb{Z}_3$--grading (see Fig.1). In both
cases the same Lie algebras arise. More precisely, there are the
following correspondences ($C_k$ and $A_{m,n}$ are symbols for Lie
algebras used in \cite{ALPW} and \cite{PSWZ}, respectively): $$
\begin{array}{lll}
C_1 \cong \mL_{0,1} \cong sl(3,\bC), \qquad & C_{10} \cong
\mL_{15,5}(1,1),\\ C_3 \cong  \mL_{12,2}(1,1), & C_{18} \cong
\mL_{18,34} \cong A_{5,33}(a,b)\oplus A_1, \\ C_7 \cong \mL_{9,1},
& C_{28} \cong \mL_{21,16} \cong A_{6,3}\oplus 2A_1, \\ C_8 \cong
\mL_{18,35},& C_{32} \cong \mL_{24,1}\cong 8A_1.\\
\end{array}
$$ \item Moreover there are still other isomorphisms between the
results of \cite{ALPW} and the present ones besides the trivial
contractions: $$
\begin{array}{lll}
C_9 \cong\mL_{20,33},& C_{30} \cong \mL_{22,2}\cong A_{3,1}\oplus
A_{3,1}\oplus 2A_1,\\ C_{27} \cong \mL_{21,12}, & C_{31} \cong
\mL_{23,3}\cong A_{3,1}\oplus 5A_1,\\ C_{29} \cong \mL_{22,3}\cong
A_{5,1}\oplus 3A_1.&\\
\end{array}
$$ The number of all contracted Lie algebras in \cite{ALPW} is 34.
Of these Lie algebras 13 algebras listed above were reobtained in
our work.

\item It also remains to investigate graded contractions of $sl(3,\C)$
starting from the fine gradings which are neither toroidal nor
Pauli \cite{HPPT}. Ultimate goal should be a comprehensive
description of all graded contractions of $sl(3,\C)$. For
convenience of the reader we reproduce the two fine gradings of
$sl(3,\bC)$ which are neither Pauli nor toroidal. They are shown
in Fig.~1, as $\Gamma_a$ and $\Gamma_d$. For more details see
\cite{HPPT}. The former grading group is $\Z_2\times \Z_2\times
\Z_2$, $$sl(3,\bC) =  L_{001}\oplus L_{111}\oplus L_{101} \oplus
L_{011}\oplus L_{110} \oplus L_{010}\oplus L_{100},\ $$ the
grading subspaces are generated by the matrices
\begin{align*}
&\left(\begin{smallmatrix}
        a&0&0\\
        0&b&0\\
        0&0&-a-b
        \end{smallmatrix}\right),\
\left(\begin{smallmatrix}
        0&1&0\\
        1&0&0\\
        0&0&0
        \end{smallmatrix}\right),\
\left(\begin{smallmatrix}
        0&0&1\\
        0&0&0\\
        1&0&0
        \end{smallmatrix}\right),\
\left(\begin{smallmatrix}
        0&0&0\\
        0&0&1\\
        0&1&0
        \end{smallmatrix}\right),\
\left(\begin{smallmatrix}
        0&-1&0\\
        1&0&0\\
        0&0&0
        \end{smallmatrix}\right),\
\left(\begin{smallmatrix}
        0&0&0\\
        0&0&-1\\
        0&1&0
        \end{smallmatrix}\right),\
\left(\begin{smallmatrix}
        0&0&-1\\
        0&0&0\\
        1&0&0
        \end{smallmatrix}\right).
\end{align*}
For the latter $ \Z_8$-grading we have $$sl(3,{C}) =  L_{0}\oplus
L_{1}\oplus L_{2} \oplus L_{3}\oplus L_{4} \oplus L_{5}\oplus
L_{6}\oplus L_{7}, $$
\begin{align*}
&\left(\begin{smallmatrix}
        0&0&0\\
        0&1&0\\
        0&0&-1
        \end{smallmatrix}\right),\
\left(\begin{smallmatrix}
        0&1&0\\
        0&0&0\\
        -1&0&0
        \end{smallmatrix}\right),\
\left(\begin{smallmatrix}
        0&0&0\\
        0&0&1\\
        0&0&0
        \end{smallmatrix}\right),\
\left(\begin{smallmatrix}
        0&0&1\\
        1&0&0\\
        0&0&0
        \end{smallmatrix}\right),\
\left(\begin{smallmatrix}
        2&0&0\\
        0&-1&0\\
        0&0&-1
        \end{smallmatrix}\right), \
\left(\begin{smallmatrix}
        0&1&0\\
        0&0&0\\
        1&0&0
        \end{smallmatrix}\right),\
\left(\begin{smallmatrix}
        0&0&0\\
        0&0&0\\
        0&1&0
        \end{smallmatrix}\right),\
\left(\begin{smallmatrix}
        0&0&1\\
        -1&0&0\\
        0&0&0
        \end{smallmatrix}\right).
\end{align*}

\item Present work raises further interesting questions.
  \begin{itemize}
  \item Why, among the contracted Lie algebras, no algebra with
  non-trivial L\'evy decomposition appears?
  \item  During the contraction the original symmetry group
of the grading in general enlarges. One can see it by comparing
initial grading symmetry and symmetry group of a contracted Lie
algebra containing Abelian algebras.
  \item One can study graded contractions preserving a
 favourite
subalgebra with the rest of the Lie algebra fine graded.
  \item  Simultaneous grading of Lie algebras and their representations
allows one to study simultaneous contractions of both: the Lie
algebra and its action on the representation space. Thus  the
representation remains representation of the contracted Lie
algebra. It was shown in \cite{MP} that the solutions found in
this paper can serve  also for describing the action of the
contracted Lie algebra on its representation space.
  \item It would be very interesting to know graded contractions of real forms of $sl(3,\C)$ for applications
in physics and elsewhere. It is conceivable that they could be
studied by splitting
 each complex solution into several which are valid for one
 or another real form.
  \end{itemize}

\end{itemize}

\ack{ Work was supported in part by the National Science and
Engineering Research Council of Canada and by the Ministry of
Education of Czech Republic (research plan MSM210000018 and by the
Czech Universities Development Fund, project Nr. 2509/2003). We
are grateful to E. Pelantov\'a and M.~Havl\'\i\v cek for numerous
stimulating and inquisitive discussions. Finally, we acknowledge
the hospitality extended to us at the Czech Technical University
in Prague (J.~P.), at the Universit\'e de Montr\'eal (J.~H., P.~N.
and J.~T.), and at the Universidad de Valladolid (J.~H., P.~N. and
J.~T.) during the work on the project.}

\section*{Appendix A}
The complete list of non-equivalent solutions of $\es$ has 188
entries. The list is divided according to the numbers of zeros
$\nu$ among 24 relevant parameters in the contraction matrices
$\ep^{\nu,i}$. Subscript $C$ or $D$ denotes continuous or discrete
solution, respectively. In the list $a\neq 0$ and $b\neq 0$ are
otherwise arbitrary complex parameters; zeros are shown as dots.

\subsection*{ Trivial solutions $\ep^{0,1}$ and  $\ep^{24,1}$} \
$$
\left(

  \right)
$$

\subsection*{Solutions with 9 zeros $\ep^{9,1}$} $$\left(
%
  \right)_{\hspace{-5pt} C}$$

\subsection*{Solutions with 12 zeros $\ep^{12,1},\ep^{12,2}$} $$\left(
%
  \right)_{\hspace{-5pt} *}$$
{\small * $\RE b >0 \vee (\RE b =0 \wedge \IM b
>0)  $; continuous for $a=1,b=1$, otherwise discrete.}
\subsection*{Solutions with 15 zeros $\ep^{15,1},\dots,\ep^{15,7}$}
$$\left(
%
  \right)_{\hspace{-5pt} D}$$
{\small ** Continuous for $a=1,b=1$, otherwise discrete.}
\subsection*{Solutions with 16 zeros $\ep^{16,1},\dots,\ep^{16,7}$} $$
\left(
%
  \right)_{\hspace{-5pt} D}$$
{\small $\dagger$ Continuous for $a=1$, otherwise discrete.}
\subsection*{Solutions with 17 zeros
$\ep^{17,1},\dots,\ep^{17,17}$} $$\left(
%
  \right)_{\hspace{-5pt} D}$$
{\small $\ddagger$ Continuous for $a=1$, otherwise discrete.}
\subsection*{Solutions with 18 zeros $\ep^{18,1},\dots,\ep^{18,36}$}
$$ \left(
%
  \right)_{\hspace{-5pt} D}$$
{\small * Continuous for $a=1$, otherwise discrete. }
\subsection*{Solutions with 19 zeros $\ep^{19,1},\dots,\ep^{19,45}$}
$$\left(
%
  \right)_{\hspace{-5pt} D}    $$

\subsection*{Solutions with 20 zeros $\ep^{20,1},\dots,\ep^{20,42}$}
$$\left(
%
  \right)_{\hspace{-5pt} D} $$

\subsection*{Solutions with 21 zeros $\ep^{21,1},\dots,\ep^{21,21}$}
$$ \left(
%
  \right)_{\hspace{-5pt} C}$$

\subsection*{Solutions with 22 zeros $\ep^{22,1},\dots,\ep^{22,7}$}
$$ \left(
%
  \right)_{\hspace{-5pt} C} $$

\subsection*{Solutions with 23 zeros $\ep^{23,1}$}
$$\left( %
 \right)_{\hspace{-5pt} C}$$

\section*{Appendix B}
The list of all contracted Lie algebras of the Pauli graded
$sl(3,\bC)$ is presented in tabular form. In the table only
non-decomposable non-Abelian parts of contracted Lie algebras are
given. (For the structure of decomposable Lie algebras see
\eqref{decom}). Algebras are divided into classes according to
dimensions of the derived series, the lower central series and the
upper central series. For each listed Lie algebra we give its
non-zero commutation relations, dimension of the algebra of
derivations ($\mathcal{D}$) and the type $T$ of contraction (C ---
continuous, D --- discrete). For solvable non-nilpotent algebras
the number $\tau$ of formal invariants and the dimension of the
radical are added. For nilpotent algebras the Casimir operators
are computed. Commutation relations are written in simplified form
without the $\omega$-coefficients, whenever this is possible. For
all Lie algebras with dimensions lower than 6 and for
6-dimensional nilpotent algebras the names from \cite{PSWZ} are
also given.


\renewcommand\arraystretch{0.5}

\begin{sidewaystable}
$$
\scriptsize
\begin{tabular}{llllllcc}

\multicolumn{8}{l}{Appendix B. Solvable Lie algebras} \\

\hline Series & Algebra & Commutation relations & $\tau$ & T & Nilradical & $\mD$ & Name\\

\hline$(430) (43) (0)$ & $\mL'_{21,9}$& $[e_1,e_4]=e_2,\ [e_2,e_4]=e_3,\ [e_3,e_4]=e_1$ &$2$& D & $3A_1$ &  6 & $A_{4,6}(\frac{\pm2}{\sqrt{3}},\frac{\mp1}{\sqrt{3}})$\\

\hline $(530) (53) (0)$ & $\mL'_{18,34}$& $[e_1,e_4]=e_2,\ [e_1,e_5]=e_3,\ [e_2,e_4]=e_3,\ [e_2,e_5]=e_1,\ [e_3,e_4]=e_1,\ [e_3,e_5]=e_2$ & $1$ & D & $3A_1$ & 6 & $A_{5,33}(-1,-1)$ \\

\hline $(640) (643) (12)$ & $\mL'_{20,16}$ & $[e_1,e_6]=e_3,\ [e_3,e_6]=e_4,\ [e_4,e_6]=e_1,\ [e_5,e_6]=e_2 $ & $4$ & D & $5A_1$ &  10 &  \\

\hline $(740) (743) (12)$ & $\mL'_{17,17}$& $[e_2,e_6]=e_3,\ [e_2,e_7]=e_4,\ [e_3,e_6]=e_4,\ [e_3,e_7]=e_2,$ & $3$ & D & $5A_1$ & 11\\

&&$[e_4,e_6]=e_2,\ [e_4,e_7]=e_3,\ [e_5,e_6]=e_1$\\

\hline $(750) (7543) (123)$ & $\mL'_{19,29}$ & $[e_1,e_7]=e_3,\ [e_2,e_7]=e_5,\ [e_4,e_7]=e_2,\ [e_5,e_7]=e_4,\ [e_6,e_7]=e_1 $ & $5$ & D & $6A_1$ &12\\

\hline $(760) (76) (0)$ & $\mL'_{18,29}(a)$& $[e_1,e_7]=e_3,\ [e_2,e_7]=-ae_5,\ [e_3,e_7]=e_6,\ [e_4,e_7]=e_2,\ [e_5,e_7]=e_4,$ & $5$ & D & $6A_1$ &12\\

&&$[e_6,e_7]=e_1,\quad 0<|a|\leq 1,\ a\neq \pm1$ \\

& &$a=1$& & C & &12\\

& &$a=-1$& & D & &18\\

\hline $(7630) (76) (0)$ & $\mL'_{15,1}$& $[e_1,e_2]=-e_5,\ [e_1,e_3]=-e_6,\ [e_1,e_7]=e_3,\ [e_2,e_3]=e_4,\ [e_2,e_7]=e_1,$ & $3$ &C & $\mL'_{21,16}$ & 9\\

&&$[e_3,e_7]=e_2,\ [e_4,e_7]=-e_6,\ [e_5,e_7]=e_4,\ [e_6,e_7]=e_5$\\

\hline $(840) (843) (13)$ & $\mL_{16,7}$& $[e_1,e_3]=-e_5,\ [e_1,e_4]=-e_8,\ [e_2,e_3]=e_7,\ [e_3,e_5]=e_8,$ & $4$ & D & $6A_1$ & 14\\

&&$[e_3,e_8]=e_1,\ [e_4,e_5]=e_1,\ [e_4,e_6]=e_7,\ [e_4,e_8]=e_5$ \\

\hline $(840) (843) (14)$ & $\mL_{19,32}$ & $[e_1,e_3]=e_5,\ [e_2,e_3]=e_7,\ [e_3,e_5]=e_8,\ [e_3,e_8]=e_1,\ [e_4,e_6]=e_7$ & $4$ & D & $\mL'_{23,1}\oplus 4A_1$ & 17\\

\hline $(850) (853) (23)$ & $\mL_{16,5}$& $[e_1,e_3]=-e_5,\ [e_1,e_4]=-e_8,\ [e_2,e_3]=e_7,\ [e_2,e_4]=e_6,\ $ & $4$ & D & $6A_1$& 13\\

&&$[e_3,e_5]=e_8,\ [e_3,e_8]=e_1,\ [e_4,e_5]=e_1,\ [e_4,e_8]=e_5$ \\

\hline $(850) (853) (24)$ & $\mL_{19,27}$ & $[e_1,e_3]=e_5,\ [e_2,e_3]=e_7,\ [e_2,e_4]=e_6,\ [e_3,e_5]=e_8,\ [e_3,e_8]=e_1$ &  $4$ & D & $\mL'_{23,1}\oplus 4A_1$ & 16\\

\hline $(850) (8543) (123)$ & $\mL_{15,6}(a)$& $[e_1,e_3]=e_5,\ [e_1,e_4]=-ae_8,\ [e_2,e_3]=-e_7,\ [e_2,e_4]=-e_6,\ [e_3,e_5]=e_8,$ & $4$ & D & $6A_1$ & 13\\

&&$[e_3,e_6]=e_2,\ [e_3,e_7]=e_6,\ [e_4,e_6]=e_7,\ [e_4,e_7]=e_2$\\

\hline $(850) (8543) (124)$ & $\mL_{18,28}$& $[e_1,e_3]=e_5,\ [e_2,e_3]=e_7,\ [e_2,e_4]=e_6,\ [e_3,e_5]=e_8,\ [e_3,e_8]=e_1,\ [e_4,e_6]=e_7$ & $4$ & D &$\mL'_{22,1}\oplus 3A_1$& 13\\

\hline $(850) (8543) (134)$ & $\mL_{18,27}$& $[e_1,e_3]=e_5,\ [e_2,e_3]=e_7,\ [e_2,e_4]=e_6,\ [e_3,e_5]=e_8,\ [e_3,e_7]=e_6,\ [e_3,e_8]=e_1$ & $4$ & D & $\mL'_{23,1}\oplus 4A_1$ & 15\\

\hline $(860) (86) (0)$ & $\mL_{12,2}(a,b)$& $[e_1,e_3]=-e_5,\ [e_1,e_4]=-a(\omega+1)e_8,\ [e_2,e_3]=-(\omega+1)e_7,\ [e_2,e_4]=-e_6,$ & $4$ & D & $6A_1$ & 12\\

&&$[e_3,e_5]=e_8,\ [e_3,e_6]=b(\omega+1)e_2,\ [e_3,e_7]=b(\omega+1)e_6,\ [e_3,e_8]=e_1,$\\

&&$[e_4,e_5]=a(\omega+1)e_1,\ [e_4,e_6]=e_7,\ [e_4,e_7]=be_2,\ [e_4,e_8]=a(\omega+1)e_5,$\\

&&$ (\mathrm{Re}\ b>0) \vee (\mathrm{Re}\ b=0 \wedge \mathrm{Im}\ b > 0)$\\

&&$ a=b=I $& & D & &18 \\

&&$a=b=1$ &  & C & &12\\

\hline $(8630) (86) (0)$ & $\mL_{9,1}$& $[e_1,e_3]=e_5,\ [e_1,e_4]=(\omega+1)e_8,\ [e_1,e_5]=e_7,\ [e_1,e_6]=(\omega+1)e_4,$ & $2$ & C & $\mL'_{21,16}$ & 9\\

&&$[e_1,e_7]=e_3,\ [e_1,e_8]=(\omega+1)e_6,\ [e_2,e_3]=(\omega+1)e_7,\ [e_2,e_4]=e_6,$\\

&&$[e_2,e_5]=(\omega+1)e_3,\ [e_2,e_6]=e_8,\ [e_2,e_7]=(\omega+1)e_5,\ [e_2,e_8]=e_4,$\\

&&$[e_4,e_6]=-e_7,\ [e_4,e_8]=-(\omega+1)e_5,\ [e_6,e_8]=-\omega e_3$\\

\hline $(8710) (87) (1)$ & $\mL_{15,4}$& $[e_1,e_3]=e_5,\ [e_1,e_6]=e_4,\ [e_2,e_6]=e_8,\ [e_2,e_7]=e_5,\ [e_3,e_6]=e_2,\ [e_4,e_6]=e_7,$ & $2$&C& $\mL'_{21,20}$&11\\

&&$[e_4,e_8]=e_5,\ [e_6,e_7]=-e_1,\ [e_6,e_8]=e_3$\\

\hline $(8740) (87) (1)$& $\mL_{12,1}$& $[e_1,e_3]=e_5,\ [e_1,e_4]=(\omega+1)e_8,\ [e_1,e_5]=e_7,\ [e_1,e_6]=(\omega+1)e_4,$ & $2$&C& $(740) (7410) (147)$ & 10\\

&&$[e_1,e_7]=e_3,\ [e_1,e_8]=(\omega+1)e_6,\ [e_3,e_6]=-(\omega+1)e_2,\ [e_4,e_6]=-e_7,$ & & & $\mD=15$\\

&&$[e_4,e_7]=-e_2,\ [e_4,e_8]=-(\omega+1)e_5,\ [e_5,e_8]=\omega e_2,\ [e_6,e_8]=-\omega e_3$\\

\hline

\end{tabular}
$$

\end{sidewaystable}

\begin{sidewaystable}

$$
\scriptsize
\begin{tabular}{llllccc}

\multicolumn{7}{l}{Nilpotent Lie algebras} \\

\hline Series & Algebra & Commutation relations & Casimir operators & T & $\mD$ & Name\\

\hline $(310) (310) (13)$ & $\mL'_{23,1}$ & $[e_2,e_3]=e_1$ & $e_1$ & C & 6 & $A_{3,1}$\\

\hline $(420) (4210) (124)$ & $\mL'_{22,1}$& $[e_1,e_4]=e_2,\ [e_3,e_4]=e_1$ & $e_2,\ e_1^2 - 2e_2e_3$ & C & 7 & $A_{4,1}$\\

\hline $(510) (510) (15)$ & $\mL'_{22,7}$& $[e_2,e_4]=e_1,\ [e_3,e_5]=e_1$ & $e_1$ & C & 15 & $A_{5,4}$\\

\hline $(520) (520) (25)$ & $\mL'_{22,3}$& $[e_3,e_4]=e_1,\ [e_3,e_5]=e_2$ & $e_1,\ e_2,\ e_2e_4-e_1e_5$ & C & 13 &$A_{5,1}$\\

\hline $(520) (5210) (135)$ & $\mL'_{21,7}$& $[e_1,e_4]=e_2,\ [e_3,e_4]=e_1,\ [e_3,e_5]=e_2$ & $e_2$ &  C & 10 & $A_{5,5}$\\

\hline $(530) (5320) (235)$ & $\mL'_{21,2}$& $[e_1,e_4]=e_2,\ [e_1,e_5]=e_3,\ [e_4,e_5]=e_1$ & $e_2,\ e_3,\ e_1^2 + 2e_2e_5 - 2e_3e_4$ & C & 10 & $A_{5,3} $\\

\hline $(620) (620) (26)$ & $\mL'_{21,19}$ & $[e_3,e_4]=e_1,\ [e_3,e_6]=e_2,\ [e_4,e_5]=e_2$ & $e_1,\ e_2$ & C & 17 & $A_{6,4}$ \\

\hline $(620) (6210) (146)$ & $\mL'_{21,21}$ & $[e_1,e_5]=e_2,\ [e_3,e_5]=e_1,\ [e_4,e_6]=e_2$ & $e_2,\ e_1^2-2e_2e_3$ & C & 14 & $A_{6,12}$ \\

\hline $(630) (630) (36)$ & $\mL'_{21,16}$ & $[e_4,e_5]=e_1,\ [e_4,e_6]=e_2,\ [e_5,e_6]=e_3 $ & $e_1,\ e_2,\ e_3,\ e_1e_6 - e_2e_5 + e_3e_4$ & C & 18 & $ A_{6,3}$ \\

\hline $(630) (6310) (136)$ & $\mL'_{20,1}$ & $[e_1,e_2]=e_4,\ [e_1,e_5]=e_3,\ [e_3,e_6]=e_4,\ [e_5,e_6]=e_2 $ & $ e_4,\ e_4e_5-e_2e_3 $ & C & 11 & $A_{6,14}(-1)$ \\

     & $\mL'_{20,2}$ & $[e_1,e_5]=e_2,\ [e_3,e_4]=e_2,\ [e_4,e_5]=e_1,\ [e_4,e_6]=e_3 $ & $e_2,\ 2e_2e_6+e_3^2$ & C & 12 & $A_{6,13}$ \\

\hline $(630) (6310) (246)$ & $\mL'_{21,1}$ & $[e_1,e_5]=e_3,\ [e_4,e_5]=e_1,\ [e_4,e_6]=e_2$ & $ e_2,\ e_3$ & C & 13 & $A_{6,7}$\\

  &   $\mL'_{21,3}$ & $[e_1,e_6]=e_3,\ [e_4,e_6]=e_1,\ [e_5,e_6]=e_2$ & $e_2,\ e_3,\ e_1^2-2e_3e_4,\ e_3e_5-e_1e_2$ & C & 15 & $A_{6,1}$\\

\hline $(630) (6320) (246)$ & $\mL'_{20,8}$ & $[e_1,e_4]=e_3,\ [e_1,e_5]=e_2,\ [e_4,e_5]=e_1,\ [e_4,e_6]=e_2 $ & $ e_2,\ e_3$ & C & 13 & $A_{6,9}$\\

\hline $(640) (64310) (1346)$ & $\mL'_{19,2}$ & $[e_2,e_6]=e_1,\ [e_3,e_5]=e_2,\ [e_3,e_6]=e_4,\ [e_4,e_5]=e_1,\ [e_5,e_6]=e_3 $ & $e_1,\ e_2e_4-e_1e_3$ & C & 10 & $A_{6,18}(-1)$\\

\hline $(710) (710) (17)$ & $\mL'_{21,20}$& $[e_2,e_4]=e_1,\ [e_3,e_6]=e_1,\ [e_5,e_7]=e_1$ & $e_1$ & C & 28\\

\hline $(720) (720) (27)$ & $\mL'_{20,39}$ & $[e_3,e_5]=e_2,\ [e_3,e_7]=e_1,\ [e_4,e_6]=e_1,\ [e_6,e_7]=e_2 $ & $e_1,\ e_2,\ e_1e_2e_7-e_1^2e_5+e_2^2e_4$ & C & 19\\

 & $\mL'_{21,11}$& $[e_3,e_5]=e_1,\ [e_4,e_6]=e_2,\ [e_4,e_7]=e_1$ & $e_1,\ e_2,\ e_2e_7-e_1e_6$ & C & 21\\

\hline $(720) (7210) (157)$ & $\mL'_{20,40}$ & $[e_1,e_5]=e_2,\ [e_3,e_5]=e_1,\ [e_3,e_6]=e_2,\ [e_4,e_7]=e_2 $ & $e_2$ & C & 19\\

\hline $(730) (730) (37)$ & $\mL'_{21,12}$& $[e_4,e_6]=e_1,\ [e_4,e_7]=e_2,\ [e_5,e_7]=e_3$ & $e_1,\ e_2,\ e_3$ & C & 20\\

\hline $(730) (7310) (137)$ & $\mL'_{19,44}$ & $[e_1,e_6]=e_2,\ [e_3,e_5]=e_2,\ [e_4,e_6]=e_1,\ [e_4,e_7]=e_2,\ [e_5,e_7]=e_3 $ & $e_2$ & D & 13\\

 & $\mL'_{20,42}$ & $[e_1,e_6]=e_3,\ [e_2,e_5]=e_3,\ [e_4,e_6]=e_1,\ [e_5,e_7]=e_2$ & $e_3,\ e_1^2-2e_3e_4,\ e_2^2+2e_3e_7$ & D & 14\\

 & $\mL'_{19,1}$ & $[e_1,e_2]=e_4,\ [e_1,e_5]=e_3,\ [e_1,e_6]=e_2,\ [e_3,e_7]=e_4,\ [e_5,e_7]=e_2 $ & $e_4,\ e_2^2-2e_4e_6,\ e_4e_5-e_2e_3$ & C & 15\\

\hline $(730) (7310) (147)$ & $\mL'_{19,7}$ & $[e_1,e_6]=e_3,\ [e_2,e_4]=e_3,\ [e_4,e_6]=e_1,\ [e_4,e_7]=e_2,\ [e_5,e_7]=e_3 $ & $e_3$ & C & 16\\

\hline $(730) (7310) (257)$ & $\mL'_{20,19}$ & $[e_1,e_6]=e_2,\ [e_4,e_6]=e_1,\ [e_4,e_7]=e_2,\ [e_5,e_7]=e_3 $ & $e_2,\ e_3,\ e_1^2e_3+2e_2^2e_5-2e_2e_3e_4$ & D & 15\\

 & $\mL'_{20,5}$ & $[e_1,e_6]=e_3,\ [e_4,e_6]=e_1,\ [e_4,e_7]=e_2,\ [e_5,e_7]=e_3 $ & $e_2,\ e_3,\ e_1^2+2e_2e_5-2e_3e_4$ & C & 17\\

 & $\mL'_{20,41}$ & $[e_1,e_6]=e_3,\ [e_4,e_6]=e_1,\ [e_5,e_6]=e_2,\ [e_5,e_7]=e_3 $ & $e_2,\ e_3,\ e_1^2-2e_3e_4$ & C & 18\\

 & $\mL'_{20,14}$ & $[e_1,e_6]=e_2,\ [e_4,e_6]=e_1,\ [e_4,e_7]=e_2,\ [e_5,e_6]=e_3 $ & $e_2,\ e_3,\ e_2e_5-e_1e_3$ & C & 19\\

\hline $(730) (7320) (257)$ & $\mL'_{19,19}$ & $[e_1,e_4]=e_3,\ [e_1,e_5]=e_2,\ [e_4,e_5]=e_1,\ [e_4,e_6]=e_2,\ [e_6,e_7]=e_3 $ & $e_2,\ e_3,\ e_1^2e_3-2e_2^2e_7-2e_2e_3e_4+2e_3^2e_5$ & D & 14 \\

 & $\mL'_{19,17}$ & $[e_1,e_4]=e_3,\ [e_1,e_6]=e_2,\ [e_4,e_6]=e_1,\ [e_4,e_7]=e_2,\ [e_5,e_6]=e_3 $ & $e_2,\ e_3,\ e_1e_2e_3-e_2^2e_5+e_3^2e_7$ & C & 16\\

 & $\mL'_{20,9}$ & $[e_1,e_4]=e_2,\ [e_1,e_6]=e_3,\ [e_4,e_6]=e_1,\ [e_5,e_7]=e_3 $ & $e_2,\ e_3,\ e_1^2+2e_2e_6-2e_3e_4$ & D & 16\\

\hline $(740) (7410) (147)$ & $\mL'_{18,13}$& $[e_1,e_6]=e_3,\ [e_2,e_5]=e_3,\ [e_4,e_7]=e_3,\ [e_5,e_6]=e_1,\ [e_5,e_7]=e_2,$ & $e_3$ & C & 15\\

&& $[e_6,e_7]=e_4$\\

\hline $(740) (7410) (247)$ & $\mL'_{19,8}$ & $[e_1,e_6]=e_3,\ [e_2,e_5]=e_3,\ [e_5,e_6]=e_1,\ [e_5,e_7]=e_2,\ [e_6,e_7]=e_4$ & $e_3,\ e_4,\ 2e_1e_4+e_2^2+2e_3e_7$ & C & 16\\

\hline $(740) (7410) (357)$ & $\mL'_{20,6}$ & $[e_1,e_6]=e_3,\ [e_5,e_6]=e_1,\ [e_5,e_7]=e_2,\ [e_6,e_7]=e_4 $ & $e_2,\ e_3,\ e_4$ & C & 18\\

\hline $(740) (7420) (247)$ & $\mL'_{18,9}$& $[e_1,e_5]=e_3,\ [e_1,e_6]=e_2,\ [e_4,e_5]=e_2,\ [e_4,e_7]=e_3,\ [e_5,e_6]=e_1,$ & $e_2,\ e_3,\ e_1^2e_3+2e_2^2e_7-2e_2e_3e_5+e_2e_4^2+2e_3^2e_6$ & D & 11\\
&& $[e_5,e_7]=e_4 $\\

 & $\mL'_{19,9}$ & $[e_1,e_6]=e_3,\ [e_2,e_5]=e_3,\ [e_2,e_7]=e_4,\ [e_5,e_6]=e_1,\ [e_5,e_7]=e_2 $ & $e_3,\ e_4,\ e_1^2e_4+e_2^2e_3+2e_3^2e_7-2e_3e_4e_5$ & D & 12\\

 & $\mL'_{20,7}$ & $[e_1,e_6]=e_3,\ [e_2,e_7]=e_4,\ [e_5,e_6]=e_1,\ [e_5,e_7]=e_2 $ & $e_3,\ e_4,\ e_1^2e_4+e_2^2e_3-2e_3e_4e_5$ & D & 13\\

 & $\mL'_{19,4}$ & $[e_1,e_2]=e_4,\ [e_1,e_5]=e_3,\ [e_3,e_5]=e_6,\ [e_3,e_7]=e_4,\ [e_5,e_7]=e_2 $ & $e_4,\ e_6,\ e_2e_3-e_4e_5+e_6e_7$ & C & 15\\

 & $\mL'_{19,5}$ & $[e_1,e_5]=e_2,\ [e_1,e_6]=e_4,\ [e_3,e_5]=e_4,\ [e_5,e_6]=e_1,\ [e_5,e_7]=e_3 $ & $e_2,\ e_4,\ e_3^2+2e_4e_7$ & C & 16\\

 & $\mL'_{20,13}$ & $[e_1,e_7]=e_3,\ [e_2,e_7]=e_4,\ [e_5,e_7]=e_1,\ [e_6,e_7]=e_2 $ & $e_3,\ e_4,\ e_1^2-2e_3e_5,\ e_2^2-2e_4e_6,\ e_2e_3-e_1e_4$ & C & 19\\

\hline $(740) (7420) (357)$ & $\mL'_{20,3}$ & $[e_1,e_5]=e_2,\ [e_1,e_6]=e_4,\ [e_5,e_6]=e_1,\ [e_5,e_7]=e_3 $ & $ e_2,\ e_3,\ e_4$ & C & 17\\

\hline $(740) (74310) (1357)$ & $\mL'_{18,1}$& $[e_1,e_7]=e_3,\ [e_2,e_5]=e_1,\ [e_2,e_7]=e_4,\ [e_4,e_5]=e_3,\ [e_5,e_6]=e_4,$ & $e_3,\ e_2e_3-e_1e_4,\ e_4^2+2e_3e_6$ & C & 13\\

&& $[e_5,e_7]=e_2$\\

\hline $(7510) (754210) (12457)$ & $\mL'_{17,1}$& $[e_1,e_2]=-e_4,\ [e_1,e_7]=e_5,\ [e_2,e_6]=e_1,\ [e_2,e_7]=e_3,$ & $e_4,\ e_5^2-2e_3e_4,\ e_2e_5-e_1e_3+e_4e_7$ & C & 12\\

&&$[e_3,e_6]=e_5,\ [e_5,e_6]=e_4,\ [e_6,e_7]=e_2$\\

\hline

\end{tabular}
$$
\end{sidewaystable}

\begin{sidewaystable}
$$
\scriptsize
\begin{tabular}{llllcc}

\hline Series & Algebra & Commutation relations & Casimir operators & T & $\mD$ \\

\hline $(820) (820) (28)$ & $\mL_{18,35}$& $[e_1,e_3]=e_5,\ [e_1,e_8]=e_6,\ [e_2,e_4]=e_6,$ & $e_5,\ e_6$ & C & 22\\

&&$[e_2,e_7]=e_5,\ [e_3,e_7]=e_6,\ [e_4,e_8]=e_5$\\

 & $\mL_{19,42}$ & $[e_1,e_3]=e_5,\ [e_1,e_8]=e_6,\ [e_2,e_4]=e_6,\ [e_2,e_7]=e_5,\ [e_3,e_7]=e_6 $ & $e_5,\ e_6$ & C & 24\\

 & $\mL_{20,32}$ & $[e_1,e_3]=e_5,\ [e_1,e_4]=e_8,\ [e_2,e_6]=e_8,\ [e_2,e_7]=e_5 $ & $e_5,\ e_8,\ e_3e_8-e_4e_5,\ e_5e_6-e_7e_8$& C & 26\\

 & $\mL_{20,38}$ & $[e_1,e_3]=e_5,\ [e_1,e_8]=e_6,\ [e_2,e_4]=e_6,\ [e_3,e_7]=e_6$ & $e_5,\ e_6 $& C & 28\\

\hline $(830) (830) (38)$ & $\mL_{19,41}$ & $[e_1,e_3]=e_5,\ [e_1,e_4]=e_8,\ [e_2,e_3]=e_7,\ [e_2,e_6]=e_8,\ [e_4,e_6]=e_7$ & $e_5,\ e_7,\ e_8,\ e_1e_7^2-e_2e_5e_7-e_3e_8^2+e_4e_5e_8+e_6e_7e_8$& C & 20\\

 & $\mL_{20,35}$ & $[e_1,e_3]=e_5,\ [e_1,e_4]=e_8,\ [e_2,e_6]=e_8,\ [e_4,e_6]=e_7 $ & $e_5,\ e_7,\ e_8,\ e_2e_5e_7+e_3e_8^2-e_4e_5e_8$&C & 23\\

 & $\mL_{20,31}$ & $[e_1,e_3]=e_5,\ [e_1,e_6]=e_4,\ [e_2,e_6]=e_8,\ [e_2,e_7]=e_5$ & $e_4,\ e_5,\ e_8,\ e_3e_4-e_5e_6+e_7e_8$ & C & 26\\

\hline $(830) (8310) (138)$ & $\mL_{17,5}$& $[e_1,e_3]=e_5,\ [e_1,e_6]=e_4,\ [e_1,e_7]=e_3,\ [e_2,e_7]=e_5,\ $ & $e_5,\ e_3e_4-e_5e_6$&D & 16\\

&&$[e_2,e_8]=e_4,\ [e_4,e_8]=e_5,\ [e_6,e_8]=e_3$\\

 & $\mL_{18,3}$& $[e_1,e_3]=e_5,\ [e_1,e_6]=e_4,\ [e_1,e_7]=e_3,$ & $e_5,\ e_3^2-2e_5e_7,\ e_3e_4-e_5e_6,\ e_4^2-2e_2e_5$&D & 19\\

&&$[e_2,e_8]=e_4,\ [e_4,e_8]=e_5,\ [e_6,e_8]=e_3$\\

\hline $(830) (8310) (148)$ & $\mL_{18,2}$& $[e_1,e_3]=e_5,\ [e_1,e_6]=e_4,\ [e_1,e_7]=e_3,$ & $e_5,\ e_3e_4-e_5e_6$& C & 18\\

&&$[e_2,e_7]=e_5,\ [e_4,e_8]=e_5,\ [e_6,e_8]=e_3$\\

\hline $(830) (8310) (158)$ & $\mL_{19,3}$ & $[e_1,e_3]=e_5,\ [e_1,e_6]=e_4,\ [e_2,e_7]=e_5,\ [e_4,e_8]=e_5,\ [e_6,e_8]=e_3 $ & $e_5,\ e_3e_4-e_5e_6$ & C & 20\\

\hline $(830) (8310) (268)$ & $\mL_{18,24}$& $[e_1,e_3]=e_5,\ [e_1,e_4]=e_8,\ [e_2,e_3]=e_7,$ & $e_7,\ e_8$ & D & 19\\

&& $[e_2,e_6]=e_8,\ [e_3,e_5]=e_8,\ [e_4,e_6]=e_7$\\

 & $\mL_{19,23}$ & $[e_1,e_3]=e_5,\ [e_1,e_4]=e_8,\ [e_2,e_6]=e_8,\ [e_3,e_5]=e_8,\ [e_4,e_6]=e_7 $ & $e_7,\ e_8$ & D & 20\\

 & $\mL_{19,24}$ & $[e_1,e_3]=e_5,\ [e_2,e_3]=e_7,\ [e_2,e_6]=e_8,\ [e_3,e_5]=e_8,\ [e_4,e_6]=e_7  $ & $e_7,\ e_8,\ 2e_1e_8+e_5^2,\ e_2e_7e_8-e_4e_8^2+e_5e_7^2$&D & 20\\

 & $\mL_{19,31}$ & $[e_1,e_3]=e_5,\ [e_1,e_4]=e_8,\ [e_2,e_3]=e_7,\ [e_3,e_5]=e_8,\ [e_4,e_6]=e_7$ & $e_7,\ e_8,\ e_2e_8+e_5e_7,\ 2e_1e_7e_8+e_5^2e_7+2e_6e_8^2$ &D & 21\\

 & $\mL_{20,11}$ & $[e_1,e_3]=e_5,\ [e_2,e_6]=e_8,\ [e_3,e_5]=e_8,\ [e_4,e_6]=e_7 $ & $e_7,\ e_8,\ e_2e_7-e_4e_8,\ 2e_1e_8+e_5^2$& D & 21\\

 & $\mL_{20,17}$ & $[e_1,e_3]=e_5,\ [e_2,e_3]=e_7,\ [e_3,e_5]=e_8,\ [e_4,e_6]=e_7 $ & $e_7,\ e_8,\ e_2e_8+e_5e_7,\ 2e_1e_8+e_5^2$&D & 23\\

 & $\mL_{19,43}$ & $[e_1,e_3]=e_5,\ [e_1,e_4]=e_8,\ [e_2,e_3]=e_7,\ [e_2,e_6]=e_8,\ [e_3,e_5]=e_8$ & $e_7,\ e_8$& C & 24\\

\hline $(830) (8320) (268)$ & $\mL_{17,14}$& $[e_1,e_3]=e_5,\ [e_1,e_4]=e_8,\ [e_1,e_5]=e_7,\ [e_2,e_3]=e_7,$ & $e_7,\ e_8$ & D & 16\\

&&$[e_2,e_6]=e_8,\ [e_3,e_5]=e_8,\ [e_4,e_6]=e_7$\\

 & $\mL_{18,20}$ & $[e_1,e_3]=e_5,\ [e_1,e_4]=e_8,\ [e_1,e_5]=e_7,$ & $e_7,\ e_8$ & D & 18\\

&&$[e_2,e_6]=e_8,\ [e_3,e_5]=e_8,\ [e_4,e_6]=e_7$\\

 & $\mL_{18,18}$& $[e_1,e_3]=e_5,\ [e_1,e_4]=e_8,\ [e_1,e_5]=e_7,$ & $e_7,\ e_8$ & D & 20\\

&&$[e_2,e_3]=e_7,\ [e_2,e_6]=e_8,\ [e_3,e_5]=e_8$\\

 & $\mL_{19,21}$ & $[e_1,e_3]=e_5,\ [e_1,e_5]=e_7,\ [e_2,e_6]=e_8,\ [e_3,e_5]=e_8,\ [e_4,e_6]=e_7$ & $e_7,\ e_8,\ e_2e_7-e_4e_8,\ 2e_1e_8-2e_3e_7+e_5^2$ & D & 20\\

 & $\mL_{19,18}$ & $[e_1,e_3]=e_5,\ [e_1,e_4]=e_8,\ [e_1,e_5]=e_7,\ [e_2,e_6]=e_8,\ [e_3,e_5]=e_8 $ & $e_7,\ e_8$ & D & 22\\

\hline $(840) (840) (48)$ & $\mL_{20,33}$ & $[e_1,e_3]=e_5,\ [e_1,e_6]=e_4,\ [e_2,e_3]=e_7,\ [e_2,e_6]=e_8$ & $e_4,\ e_5,\ e_7,\ e_8$ & C & 24\\

\hline $(840) (8410) (248)$ & $\mL_{18,36}$& $[e_1,e_3]=e_5,\ [e_1,e_4]=e_8,\ [e_2,e_3]=e_7,\ $ & $e_7,\ e_8$ & D & 18\\

&&$[e_2,e_4]=e_6,\ [e_2,e_6]=e_8,\ [e_3,e_5]=e_8$\\

& $\mL_{19,45}$& $[e_1,e_3]=e_5,\ [e_2,e_3]=e_7,\ [e_2,e_4]=e_6,\ [e_2,e_6]=e_8,\ [e_3,e_5]=e_8$ & $e_7,\ e_8,\ e_5^2+2e_1e_8,\ e_6^2-2e_4e_8$ & D & 19\\

\hline $(840) (8410) (258)$ & $\mL_{18,10}$& $[e_1,e_3]=e_5,\ [e_1,e_4]=e_8,\ [e_1,e_6]=e_4,\ $ & $e_7,\ e_8$ & D & 19\\

&&$[e_2,e_3]=e_7,\ [e_2,e_6]=e_8,\ [e_3,e_5]=e_8$\\

 & $\mL_{19,6}$ & $[e_1,e_3]=e_5,\ [e_1,e_4]=e_8,\ [e_1,e_6]=e_4,\ [e_2,e_3]=e_7,\ [e_3,e_5]=e_8$ & $e_7,\ e_8,\ e_4^2-2e_6e_8,\ e_2e_8+e_5e_7$ & D & 20\\

\hline $(840) (8410) (368)$ & $\mL_{19,25}$ & $[e_1,e_3]=e_5,\ [e_1,e_4]=e_8,\ [e_2,e_3]=e_7,\ [e_2,e_4]=e_6,\ [e_3,e_5]=e_8$ & $e_6,\ e_7,\ e_8,\ 2e_1e_6e_8-2e_2e_8^2+e_5^2e_6-2e_5e_7e_8$ & D & 20\\

 & $\mL_{19,13}$ & $[e_1,e_3]=e_5,\ [e_1,e_6]=e_4,\ [e_2,e_3]=e_7,\ [e_2,e_6]=e_8,\ [e_3,e_5]=e_8$ & $e_4,\ e_7,\ e_8,\ 2e_1e_8^2-2e_2e_4e_8-2e_4e_5e_7+e_5^2e_8$&C & 21\\

 & $\mL_{20,12}$ & $[e_1,e_3]=e_5,\ [e_2,e_3]=e_7,\ [e_2,e_4]=e_6,\ [e_3,e_5]=e_8 $ & $e_6,\ e_7,\ e_8,\ 2e_1e_8+e_5^2$ & D & 22\\

 & $\mL_{20,4}$ & $[e_1,e_3]=e_5,\ [e_1,e_6]=e_4,\ [e_2,e_3]=e_7,\ [e_3,e_5]=e_8 $ & $e_4,\ e_7,\ e_8,\ e_2e_8+e_5e_7$&C & 23\\

\hline $(840) (8420) (248)$ & $\mL_{16,4}$& $[e_1,e_3]=e_5,\ [e_1,e_4]=e_8,\ [e_1,e_5]=e_7,\ [e_2,e_3]=e_7,\ $ & $e_7,\ e_8$ & D & 12\\

&&$[e_2,e_4]=e_6,\ [e_2,e_6]=e_8,\ [e_3,e_5]=e_8,\ [e_4,e_6]=e_7$\\

 & $\mL_{17,16}$& $[e_1,e_3]=e_5,\ [e_1,e_4]=e_8,\ [e_1,e_5]=e_7,\ [e_2,e_4]=e_6,\ $ & $e_7,\ e_8$ & D & 14\\

&&$[e_2,e_6]=e_8,\ [e_3,e_5]=e_8,\ [e_4,e_6]=e_7$\\

\hline

\end{tabular}
$$
\end{sidewaystable}

\begin{sidewaystable}
$$
\scriptsize
\begin{tabular}{llllcc}

\hline Series & Algebra & Commutation relations & Casimir operators & T & $\mD$ \\

\hline

 & $\mL_{17,15}$& $[e_1,e_3]=e_5,\ [e_1,e_4]=e_8,\ [e_1,e_5]=e_7,\ [e_2,e_3]=e_7, $ & $e_7,\ e_8$ & D & 15\\

&&$[e_2,e_4]=e_6,\ [e_2,e_6]=e_8,\ [e_3,e_5]=e_8$\\

 & $\mL_{18,21}$& $[e_1,e_3]=e_5,\ [e_1,e_4]=e_8,\ [e_1,e_5]=e_7,\ [e_2,e_4]=e_6, $ & $e_7,\ e_8$ & D & 16\\

&&$[e_2,e_6]=e_8,\ [e_3,e_5]=e_8$\\

 & $\mL_{18,22}$& $[e_1,e_3]=e_5,\ [e_1,e_4]=e_8,\ [e_1,e_5]=e_7,\ [e_2,e_4]=e_6,$ & $e_7,\ e_8,\ 2e_2e_7+e_6^2,\ 2e_1e_7e_8-2e_3e_7^2+e_5^2e_7+2e_6e_8^2$&D & 16\\

&&$[e_3,e_5]=e_8,\ [e_4,e_6]=e_7$\\

 & $\mL_{18,23}$& $[e_1,e_3]=e_5,\ [e_1,e_5]=e_7,\ [e_2,e_4]=e_6,\ [e_2,e_6]=e_8,\ $ & $e_7,\ e_8,\ 2e_1e_8-2e_3e_7+e_5^2,\ 2e_2e_7-2e_4e_8+e_6^2   $&D & 16\\

&&$[e_3,e_5]=e_8,\ [e_4,e_6]=e_7$\\

 & $\mL_{18,26}$& $[e_1,e_3]=e_5,\ [e_1,e_4]=e_8,\ [e_2,e_3]=e_7,\ [e_2,e_4]=e_6,$ & $e_7,\ e_8,\ 2e_1e_7e_8+e_5^2e_7+2e_6e_8^2,\ 2e_2e_7e_8+2e_5e_7^2+e_6^2e_8$ & D & 16\\

    &&$[e_3,e_5]=e_8,\ [e_4,e_6]=e_7$\\

 & $\mL_{17,2}$& $[e_1,e_3]=e_5,\ [e_1,e_6]=e_4,\ [e_1,e_7]=e_3,\ [e_3,e_6]=e_2,$ & $e_2,\ e_5,\ e_1e_2-e_3e_4+e_5e_6,\ 2e_2e_8+e_3^2-2e_5e_7$ & C & 17\\

&&$[e_4,e_7]=e_2,\ [e_4,e_8]=e_5,\ [e_6,e_8]=e_3$\\

 & $\mL_{19,22}$ & $[e_1,e_3]=e_5,\ [e_1,e_5]=e_7,\ [e_2,e_4]=e_6,\ [e_2,e_6]=e_8,\ [e_3,e_5]=e_8 $ & $e_7,\ e_8,\ 2e_4e_8-e_6^2,\ 2e_1e_8-2e_3e_7+e_5^2$& D & 17\\

 & $\mL_{19,28}$ & $[e_1,e_3]=e_5,\ [e_2,e_3]=e_7,\ [e_2,e_4]=e_6,\ [e_3,e_5]=e_8,\ [e_4,e_6]=e_7 $ & $e_7,\ e_8,\ 2e_1e_8+e_5^2,\ 2e_2e_7e_8+2e_5e_7^2+e_6^2e_8$&D & 17\\

\hline $(840) (8420) (258)$ & $\mL_{16,3}(a)$& $[e_1,e_3]=-ae_5,\ [e_1,e_4]=e_8,\ [e_1,e_5]=e_7,\ [e_1,e_6]=e_4,$ & $e_7,\ e_8$ & D & 16\\

&&$[e_2,e_3]=e_7,\ [e_2,e_6]=e_8,\ [e_3,e_5]=e_8,\ [e_4,e_6]=e_7,$\\

&&$0 < |a| \leq 1$\\

 & $\mL_{17,9}$& $[e_1,e_3]=e_5,\ [e_1,e_4]=e_8,\ [e_1,e_5]=e_7,\ [e_1,e_6]=e_4,$ & $e_7,\ e_8$ & D & 16\\

&&$[e_2,e_3]=e_7,\ [e_3,e_5]=e_8,\ [e_4,e_6]=e_7$\\

 & $\mL_{17,12}$& $[e_1,e_3]=e_5,\ [e_1,e_4]=e_8,\ [e_1,e_6]=e_4,\ [e_2,e_3]=e_7,$ & $e_7,\ e_8$ & D & 16\\

&&$[e_2,e_6]=e_8,\ [e_3,e_5]=e_8,\ [e_4,e_6]=e_7$\\

 & $\mL_{18,12}$& $[e_1,e_3]=e_5,\ [e_1,e_4]=e_8,\ [e_1,e_6]=e_4,\ [e_2,e_3]=e_7,\ $ & $e_7,\ e_8,\ e_2e_8+e_5e_7,\ 2e_1e_7e_8-e_4^2e_8+e_5^2e_7+2e_6e_8^2$&D & 17\\

&&$[e_3,e_5]=e_8,\ [e_4,e_6]=e_7$\\

 & $\mL_{18,14}$& $[e_1,e_3]=e_5,\ [e_1,e_4]=e_8,\ [e_1,e_6]=e_4,\ [e_2,e_6]=e_8,$ & $e_7,\ e_8$ & D & 17\\

&&$[e_3,e_5]=e_8,\ [e_4,e_6]=e_7$\\

 & $\mL_{18,17}$& $[e_1,e_3]=e_5,\ [e_1,e_6]=e_4,\ [e_2,e_3]=e_7,\ [e_2,e_6]=e_8,$ & $e_7,\ e_8,\ 2e_1e_7e_8-e_4^2e_8+e_5^2e_7,\ e_2e_7e_8-e_4e_8^2+e_5e_7^2$ & D & 17\\

&&$[e_3,e_5]=e_8,\ [e_4,e_6]=e_7$\\

 & $\mL_{19,15}$ & $[e_1,e_3]=e_5,\ [e_1,e_6]=e_4,\ [e_2,e_3]=e_7,\ [e_3,e_5]=e_8,\ [e_4,e_6]=e_7$ & $e_7,\ e_8,\ e_2e_8+e_5e_7,\ 2e_1e_7e_8-e_4^2e_8+e_5^2e_7$& D & 18\\

 & $\mL_{17,7}(a)$& $[e_1,e_3]=-ae_5,\ [e_1,e_4]=e_8,\ [e_1,e_5]=e_7,\ [e_1,e_6]=e_4,$ & $e_7,\ e_8$ &D & 19 \\

&&$[e_2,e_3]=e_7,\ [e_2,e_6]=e_8,\ [e_3,e_5]=e_8,\quad 0\neq a \neq 1$& \\

&&$a=1$& -//-,-//- & C & 19\\

 & $\mL_{18,6}$& $[e_1,e_3]=e_5,\ [e_1,e_4]=e_8,\ [e_1,e_5]=e_7,\ [e_1,e_6]=e_4,$ & $e_7,\ e_8,\ e_4^2-2e_6e_8,\ e_2e_8^2-e_4e_7^2+e_5e_7e_8 $& D & 19\\

&&$[e_2,e_3]=e_7,\ [e_3,e_5]=e_8$\\

 & $\mL_{18,7}$& $[e_1,e_3]=e_5,\ [e_1,e_4]=e_8,\ [e_1,e_5]=e_7,\ [e_1,e_6]=e_4,$ & $e_7,\ e_8$ & D & 20\\

&&$[e_2,e_6]=e_8,\ [e_3,e_5]=e_8$\\

 & $\mL_{18,25}(a)$& $[e_1,e_3]=e_5,\ [e_1,e_4]=-ae_8,\ [e_2,e_3]=e_7,\ [e_2,e_4]=e_6,\ $ & $e_6,\ e_8,\ e_5e_6-e_7e_8,$\\

&&$[e_3,e_5]=e_8,\ [e_3,e_7]=e_6,\quad 0 < |a| \leq 1,\ a\neq \pm1$ & $2e_1e_6e_8+2ae_2e_8^2+(1-a)e_5^2e_6+2ae_5e_7e_8$ & D & 20\\

&& $a=-1$ & -//-,-//-,-//-,-//- & D & 22\\

&& $a=1$ &-//-,-//-,-//-, $e_1e_6+e_2e_8+e_5e_7$ & C & 20\\

&& $a=0$ &-//-,-//-,-//-, $2e_1e_8+e_5^2$ & D & 21\\

\hline $(840) (8420) (368)$ & $\mL_{18,19}$& $[e_1,e_3]=e_5,\ [e_1,e_4]=e_8,\ [e_1,e_5]=e_7,\ [e_2,e_3]=e_7,$ & $e_6,\ e_7,\ e_8,$ & D & 16\\

&& $[e_2,e_4]=e_6,\ [e_3,e_5]=e_8$ & $2e_1e_6e_8-2e_2e_8^2-2e_3e_6e_7+2e_4e_7^2+e_5^2e_6-2e_5e_7e_8$\\

 & $\mL_{18,16}$& $[e_1,e_3]=e_5,\ [e_1,e_5]=e_7,\ [e_1,e_6]=e_4,\ [e_2,e_3]=e_7,$ & $e_4,\ e_7,\ e_8,$ & D & 19\\

&&$[e_2,e_6]=e_8,\ [e_3,e_5]=e_8$ & $2e_1e_8^2-2e_2e_4e_8-2e_3e_7e_8-2e_4e_5e_7+e_5^2e_8+2e_6e_7^2$\\

\hline

\end{tabular}
$$
\end{sidewaystable}

\begin{sidewaystable}

$$
\scriptsize
\begin{tabular}{llllcc}

\hline Series & Algebra & Commutation relations & Casimir operators & T & $\mD$ \\

\hline

 & $\mL_{19,20}$ & $[e_1,e_3]=e_5,\ [e_1,e_4]=e_8,\ [e_1,e_5]=e_7,\ [e_2,e_4]=e_6,\ [e_3,e_5]=e_8 $ & $e_6,\ e_7,\ e_8,\ 2e_1e_6e_8-2e_2e_8^2-2e_3e_6e_7+e_5^2e_6$ & D & 19\\

 & $\mL_{19,10}$ & $[e_1,e_3]=e_5,\ [e_1,e_5]=e_7,\ [e_1,e_6]=e_4,\ [e_2,e_3]=e_7,\ [e_3,e_5]=e_8$ & $e_4,\ e_7,\ e_8,\ e_2e_4e_8+e_4e_5e_7-e_6e_7^2$ & C & 21\\

 & $\mL_{19,11}$ & $[e_1,e_3]=e_5,\ [e_1,e_5]=e_7,\ [e_1,e_6]=e_4,\ [e_2,e_6]=e_8,\ [e_3,e_5]=e_8$ & $e_4,\ e_7,\ e_8,\ 2e_1e_8-2e_2e_4-2e_3e_7+e_5^2$ & D & 21\\

\hline $(840) (84310) (1368)$ & $\mL_{16,1}(a)$& $[e_1,e_3]=e_5,\ [e_1,e_6]=e_4,\ [e_1,e_7]=e_3,\ [e_1,e_8]=e_6,\ [e_2,e_7]=e_5 $ & $e_5,\ e_3e_4-e_5e_6$ &D & 16 \\

&&$[e_2,e_8]=-ae_4,\ [e_4,e_8]=e_5,\ [e_6,e_8]=e_3,\ 0\neq a\neq 1$\\

&&$a=1$& -//-,-//- &C &16\\

 & $\mL_{17,4}$& $[e_1,e_3]=e_5,\ [e_1,e_6]=e_4,\ [e_1,e_7]=e_3,\ [e_1,e_8]=e_6,$ & $e_5,\ e_3^2-2e_5e_7,\ e_3e_4-e_5e_6,\ 2e_2e_5-e_4^2$ & D & 16\\

&&$[e_2,e_8]=e_4,\ [e_4,e_8]=e_5,\ [e_6,e_8]=e_3$\\

\hline $(840) (84310) (1468)$ & $\mL_{17,3}$& $[e_1,e_3]=e_5,\ [e_1,e_6]=e_4,\ [e_1,e_7]=e_3,\ [e_1,e_8]=e_6,$ & $e_5,\ e_3e_4-e_5e_6$ & D & 17\\

&&$[e_2,e_7]=e_5,\ [e_4,e_8]=e_5,\ [e_6,e_8]=e_3$\\

\hline $(840) (84310) (1568)$ & $\mL_{18,4}$& $[e_1,e_3]=e_5,\ [e_1,e_6]=e_4,\ [e_1,e_8]=e_6,\ [e_2,e_7]=e_5,$ & $e_5,\ e_3e_4-e_5e_6$ & D & 19\\

&&$[e_4,e_8]=e_5,\ [e_6,e_8]=e_3$\\

\hline $(850) (8520) (258)$ & $\mL_{15,5}(a,b)$& $[e_1,e_3]=-ae_5,\ [e_1,e_4]=e_8,\ [e_1,e_5]=e_7,\ [e_1,e_6]=e_4,$ & $e_7,\ e_8$ &D & 16 \\

&&$[e_2,e_3]=e_7,\ [e_2,e_6]=e_8,\ [e_3,e_5]=-be_8,\ [e_3,e_6]=e_2,$\\

&&$[e_4,e_6]=e_7, \quad  a\neq 0$\\

&&$a=-1,\ b=0$ & -//-,-//- & D & 17\\

&&$a=1,\ b=1$ & -//-,-//- & C & 18\\

 & $\mL_{17,8}$& $[e_1,e_3]=e_5,\ [e_1,e_4]=e_8,\ [e_1,e_5]=e_7,\ [e_1,e_6]=e_4,\ $ & $e_7,\ e_8,\ e_2e_8^2-e_4e_7^2+e_5e_7e_8,\ e_2^2e_8-e_4^2e_7+2e_6e_7e_8$& D & 16\\

&&$[e_2,e_3]=e_7,\ [e_3,e_5]=e_8,\ [e_3,e_6]=e_2$\\

 & $\mL_{17,10}$& $[e_1,e_3]=e_5,\ [e_1,e_4]=e_8,\ [e_1,e_5]=e_7,\ [e_1,e_6]=e_4,$ & $e_7,\ e_8$ & D & 16\\

&&$[e_2,e_6]=e_8,\ [e_3,e_5]=e_8,\ [e_3,e_6]=e_2$\\

 & $\mL_{17,13}(a)$& $[e_1,e_3]=ae_5,\ [e_1,e_4]=e_8,\ [e_1,e_6]=e_4,\ [e_2,e_3]=e_7,$ & $e_7,\ e_8,\ e_2e_8+e_5e_7,$ & D & 17\\

&&$[e_3,e_5]=e_8,\ [e_3,e_6]=e_2,\ [e_4,e_6]=e_7,\ 0 < |a| \leq 1,\ a\neq \pm1$ & $2e_1e_7^2+(1-a)e_2^2e_8+-2ae_2e_5e_7+e_4^2e_7-2e_6e_7e_8$\\

&&$a=-1$ & -//-,-//-,-//-,-//-& D & 19\\

&&$a=1$ &  -//-,-//-,-//-, $2e_1e_7-2e_2e_5-e_4^2+2e_6e_8$ & D & 17\\

 & $\mL_{18,11}$& $[e_1,e_3]=e_5,\ [e_1,e_4]=e_8,\ [e_1,e_6]=e_4,\ [e_2,e_3]=e_7,\ $ & $e_7,\ e_8,\ e_2e_8+e_5e_7,\ e_2^2e_8-e_4^2e_7+2e_6e_7e_8$ & D & 17\\

&&$[e_3,e_5]=e_8,\ [e_3,e_6]=e_2$ \\

\hline $(850) (8520) (358)$ & $\mL_{17,11}$& $[e_1,e_3]=e_5,\ [e_1,e_4]=e_8,\ [e_1,e_5]=e_7,\ [e_1,e_6]=e_4,$ & $e_2,\ e_7,\ e_8,$ & D & 16\\

&&$[e_3,e_5]=e_8,\ [e_3,e_6]=e_2,\ [e_4,e_6]=e_7$ & $2e_1e_7e_8+2e_2e_4e_7-2e_2e_5e_8-2e_3e_7^2-e_4^2e_8+e_5^2e_7+2e_6e_8^2$\\

 & $\mL_{18,15}$& $[e_1,e_3]=e_5,\ [e_1,e_4]=e_8,\ [e_1,e_6]=e_4,\ [e_3,e_5]=e_8,$ & $e_2,\ e_7,\ e_8,\ 2e_1e_7e_8-2e_2e_5e_8-e_4^2e_8+e_5^2e_7 +2e_6e_8^2$ & D & 17\\

&&$[e_3,e_6]=e_2,\ [e_4,e_6]=e_7$ \\

 & $\mL_{19,16}$ & $[e_1,e_3]=e_5,\ [e_1,e_6]=e_4,\ [e_3,e_5]=e_8,\ [e_3,e_6]=e_2,\ [e_4,e_6]=e_7$ & $e_2,\ e_7,\ e_8,\ 2e_1e_7e_8-e_4^2e_8+e_5^2e_7$ & D & 18\\

 & $\mL_{18,8}$& $[e_1,e_3]=e_5,\ [e_1,e_4]=e_8,\ [e_1,e_5]=e_7,\ [e_1,e_6]=e_4, $ & $e_2,\ e_7,\ e_8,\ 2e_2e_4e_7-2e_2e_5e_8-e_4^2e_8+2e_6e_8^2$ & C & 20\\

&&$[e_3,e_5]=e_8,\ [e_3,e_6]=e_2$ \\

 & $\mL_{19,14}$ & $[e_1,e_3]=e_5,\ [e_1,e_6]=e_4,\ [e_2,e_3]=e_7,\ [e_3,e_5]=e_8,\ [e_3,e_6]=e_2$ & $e_4,\ e_7,\ e_8,\ e_2e_8+e_5e_7$ & C & 22\\

\hline $(850) (8520) (468)$ & $\mL_{19,12}$ & $[e_1,e_3]=e_5,\ [e_1,e_5]=e_7,\ [e_1,e_6]=e_4,\ [e_3,e_5]=e_8,\ [e_3,e_6]=e_2$ & $e_2,\ e_4,\ e_7,\ e_8$ & C & 22\\

\hline $(850) (8530) (358)$ & $\mL_{18,5}$& $[e_1,e_3]=e_5,\ [e_1,e_6]=e_4,\ [e_3,e_6]=e_2,\ [e_4,e_6]=e_7,$ & $e_2,\ e_5,\ e_7,\ e_1e_2-e_3e_4+e_5e_6-e_7e_8$ & C & 19\\

&&$[e_4,e_8]=e_5,\ [e_6,e_8]=e_3$\\

\hline $(850) (85310) (1358)$ & $\mL_{17,6}$& $[e_1,e_3]=e_5,\ [e_1,e_6]=e_4,\ [e_2,e_6]=e_8,\ [e_2,e_7]=e_5,$ & $e_5,\ e_3e_4-e_5e_6+e_7e_8$ & C & 16\\

&&$[e_4,e_6]=e_7,\ [e_4,e_8]=e_5,\ [e_6,e_8]=e_3$\\

\hline $(8510) (854210) (12468)$ & $\mL_{15,2}$& $[e_1,e_3]=e_5,\ [e_1,e_4]=e_8,\ [e_1,e_6]=e_4,\ [e_1,e_7]=-e_3, $ & $e_5,\ e_3e_4-e_5e_6+e_7e_8$ & C & 13\\

&&$[e_2,e_6]=e_8,\ [e_2,e_7]=e_5,\ [e_4,e_6]=e_7,\ [e_4,e_8]=e_5,\ [e_6,e_8]=e_3$\\

\hline $(8620) (865320) (23568)$ & $\mL_{15,3}$& $[e_1,e_3]=e_5,\ [e_1,e_4]=e_8,\ [e_1,e_6]=e_4,\ [e_1,e_7]=e_3,\ [e_3,e_6]=e_2,$ & $e_2,\ e_5,\ 2e_2e_8+e_3^2-2e_5e_7,\ e_1e_2-e_3e_4+e_5e_6+e_7e_8$ & C & 14\\

&&$[e_4,e_6]=-e_7,\ [e_4,e_7]=e_2,\ [e_4,e_8]=e_5,\ [e_6,e_8]=e_3$\\

\hline

\end{tabular}
$$
\end{sidewaystable}


\end{document}